\documentclass[apj, twocolumn]{openjournal}

\usepackage[utf8]{inputenc}

\usepackage{amsmath,amssymb,amstext}
\usepackage[breaklinks,colorlinks,
   urlcolor=blue,citecolor=blue,linkcolor=blue]{hyperref}
\usepackage{url}
\usepackage{orcidlink}

\newcommand{\isot}[2]{$^{#2}\mathrm{#1}$}
\newcommand{\isotm}[2]{{}^{#2}\mathrm{#1}}
\newcommand{\Xisot}[2]{$X(\isotm{#1}{#2})$}
\newcommand{\Xisotm}[2]{X (\isotm{#1}{#2})}
\newcommand{\Rconv}{\ensuremath{R_{\mathrm{conv}}}}
\newcommand{\Mconv}{\ensuremath{M_{\mathrm{conv}}}}
\newcommand{\Msun}{\ensuremath{\mathrm{M}_\odot}}
\newcommand{\Tauconvm}{\tau_{\mathrm{conv}}}

\newcommand{\vrms}{\ensuremath{U_{\mathrm{rms}}}}
\newcommand{\vrmsr}{\ensuremath{U_{\mathrm{rms}}(r)}}

\newcommand{\unitstyle}{\mathrm}
\newcommand{\gcc}{\unitstyle{g~cm^{-3}}} %grams per cubic centimeter
\newcommand{\maestro}{{\sffamily MAESTROeX}}

\newcommand{\microphysics}{{\sffamily Microphysics}}
\newcommand{\pynucastro}{{\sffamily pynucastro}}
\newcommand{\amrex}{{\sffamily AMReX}}
\newcommand{\yt}{{\sffamily yt}}

\journalinfo{The Open Journal of Astrophysics}

\begin{document}

\title{Simulating the Convective Urca Process with Multiple Urca Pairs in a Simmering White Dwarf}

\author{
	Brendan Boyd$^{1, 2}$ \orcidlink{0000-0002-5419-9751}}
\author{
	Ferran Poca-Amor\'{o}s$^{3}$ \orcidlink{0009-0000-1964-7734}}
\author{
	Alan Calder$^{1, 2}$ \orcidlink{0000-0001-5525-089X}}
\author{
	Dean M.\ Townsley$^{4}$ \orcidlink{0000-0002-9538-5948}}

 \affiliation{$^{1}$Department of Physics and Astronomy, 
 Stony Brook University, Stony Brook, NY 11794-3800, USA}
 \affiliation{$^{2}$Institute for Advanced Computational Science,
 Stony Brook University, Stony Brook, NY 11794-5250, USA}
 \affiliation{$^{3}$Universitat Polit\`{e}cnica de Catalunya, Barcelona, Spain}
 \affiliation{$^{4}$Department of Physics and Astronomy, University of Alabama, Tuscaloosa, AL 35487-0324, USA}

%\correspondingauthor{Brendan Boyd}
\email{boyd.brendan@stonybrook.edu}

\begin{abstract}
	Type Ia supernovae are bright thermonuclear explosions of one or more white dwarf stars.
	The exact origin and explosion mechanism for these supernovae is still poorly understood.
	In the near-Chandrasekhar mass progenitor model, a simmering phase precedes the explosion.
	During this simmering phase, central carbon burning heats the core and drives convection.
	A poorly understood aspect of this phase is the convective Urca process, a linking of weak nuclear reactions and convective mixing.
	Convective Urca has the potential to alter characteristics of the convection zone and thus alter the evolution of the white dwarf. 
	To study the convective Urca process, we use the low Mach number hydrodynamic code MAESTROeX to run 3D simulations of the convection zone.
	We build off previous work to implement a more comprehensive carbon burning network and include the A=21, A=23, and A=25 Urca pairs in the simulations.
	We compare simulations with and without the convective Urca process to isolate the direct effects the process has on the convection zone.
	We find the convective Urca process reduces the efficiency of convective mixing near the the convective boundary, but does not restrict the size of the convection zone.
	We additionally find the A=23 Urca pair to be the most important Urca pair to the convective Urca process in these simulations.
	All together, our results better inform our understanding of this complex phenomena as well as demonstrates the range of potential convective structures, particularly at the convective boundary, of a simmering white dwarf. 
\end{abstract}


\maketitle
\section{Introduction}\label{sec:intro}
    Type Ia supernovae (SNe Ia) are extremely bright events and are widely understood to be thermonuclear explosions of roughly a solar mass of degenerate white dwarf material.
	The luminosity of SNe Ia is primarily powered by the decay of \isot{Ni}{56}, which enables the light curves to be standardized \citep{phillips1993}. 
	The standardized nature, along with high luminosity, make SNe Ia well suited as standard candles and highly important to cosmology research \citep{riess1998, perlmutter1999}. 
	The progenitor system and explosion mechanism of SNe Ia are still not well understood and are important to understanding SNe Ia from first principles. 
	There have been a number of models presented to explain SN Ia which vary in both progenitor system and explosion mechanism~\citep{maoz2014, liu2023}.
	In terms of the progenitor system, one of the main classification is sub-Chandrasekhar mass and near-Chandrasekhar mass white dwarfs.
	In addition to explaining ``normal" SN Ia, there is also a need to explain the diversity of SNe Ia~\citep{taubenberger2017}, such as the relatively dim class SN Iax~\citep{foley2013}.
	We focus this paper on the near-Chandrasekhar mass progenitor model where a white dwarf gains mass via accretion from a companion star.
	As the white dwarf nears the Chandrasekhar mass limit, runaway carbon burning ignites in the core, producing the explosion~\citep{nomoto1984, woosley1986}. 

	Prior to the carbon burning flame that incinerates some or all of the white dwarf, there is a roughly thousand year long period of slow carbon burning called the simmering phase.
	During this period, central carbon burning heats the core and drives convection in the interior of the star.
	As a whole, the simmering phase can alter the structure and composition of the white dwarf~\citep{chamulak2008, piro2008}, potentially impacting the final nucleosynthesis and explosion~\citep{Umeda1999, timmes2003}.
	A poorly understood component during this simmering phase is the convective Urca process~\citep{paczynski1972}.
	Properly understanding the convective Urca process is important for 1D stellar evolution studies that model the full simmering phase~\citep{denissenkov2015, schwab2017a, piersanti2022}

	The convective Urca process links weak Urca reactions and convective mixing.
	The Urca process itself is the relation between two isotopes, called an Urca pair, linked by a beta-decay and electron capture.
	In the highly degenerate conditions of a white dwarf, these weak reactions are highly density dependent~\citep{suzuki2016}.
	This dependence creates the conditions such that the electron capture rate is favored at high densities, near the center of the star, while the beta-decays are favored at lower densities, away from the center of the star.
	If convective mixing is present, it can move material back and forth across this density gradient, leading to repeated electron captures and beta-decays.

	The primary method of modeling the simmering phase is through 1D stellar evolution models~\citep[e.g.][]{martinez-rodriguez2016, piersanti2017, schwab2017a, piersanti2022}.
	These models however must make approximations on the convective mixing, which makes studying the convective Urca process particularly challenging.
	Some work has been made to address this via a two-stream analytical approach~\citep{lesaffre2005} but is numerically difficult to implement in modern stellar evolution codes.
	The first multi-D simulations investigating convective Urca used a 2D wedge geometry with boosted reaction rates~\citep{stein-wheeler2006}.
	This 2D numerical work, and previous theoretical work~\citep{sbw1999, lesaffre2005}, indicates the convective Urca process can hinder or restrict convection during the simmering phase.
	More recent work to model the full 3D turbulent convection zone during a period of the simmering phase indicates the effects of the convective Urca process may be less significant~\citep{boyd2025, poca2026}, which we will refer to as B2025 and PA2026 respectively.
	A major drawback in these 3D simulations is the simple carbon burning network that does not accurately model the energy generation rate or the compositional changes, particularly to the neutron excess.
	The work presented here builds off the previous 3D studies with a more comprehensive reaction network that accounts for these shortcomings.
	
	In Section \ref{sec:urca}, we describe the convective Urca process in further detail including how the process can affect convection. 
	Section \ref{sec:sim_details} describes the \maestro\ code and numerical methods in our simulations. 
	In Section \ref{sec:network}, we describe the reaction network, particularly in relation to previous work.
    Then, in Section \ref{sec:init_models}, we describe the initial model with an emphasis on the chosen temperature gradient and initial composition.
    In Section \ref{sec:analysis}, we present the simulations after they have settled into a steady state and characterize important features of the convection zone. 
    We then discuss, in Section \ref{sec:discussion}, the effects of the convective Urca process on the simulation and particularly the convective boundary.
    And finally, in Section \ref{sec:conclusion}, we draw our conclusions and point to future work and questions.

    \section{Convective Urca Process} \label{sec:urca}
    The Urca process links two isotopes, called an Urca pair, via a beta-decay and electron capture reaction.
    The reactions work as follows for a pair of nuclei with the same atomic mass number, $A$, and proton numbers, $Z-1$ and $Z$ respectively:
    \begin{equation}
        \begin{split}
            \left(Z-1,A\right)           &\longrightarrow \: \left(Z,A\right) + e^{-} + \bar{\nu}_e  \\
            \left(Z,A\right) + e^{-} &\longrightarrow \: \left(Z-1,A\right) + \nu_e
        \end{split}
    \end{equation}
	For both reactions, the emitted neutrinos free-stream from the white dwarf.
    In a degenerate white dwarf, these weak reactions depend primarily on the electron density. 
    In most regions only one reaction will be active, i.e.\ electron captures at higher densities and beta-decays at lower densities. 
    The transition between these regions, where the reaction rates are equal, is called the Urca shell.
    The convective Urca process is the process in which convection transports material back and forth across the Urca shell.
    As convection cyclically mixes material, the Urca process will occur continuously. 
    The cyclical nature enables the convective Urca process to have a meaningful impact even with relatively small abundances of an Urca pair. 

    There are a number of Urca pairs relevant to simmering white dwarfs, but we focus this paper on \isot{Ne}{21} -- \isot{F}{21}, \isot{Na}{23} -- \isot{Ne}{23}, and \isot{Mg}{25} -- \isot{Na}{25}, which we denote as the A=21, A=23, and A=25 Urca pairs, respectively. 
    The A=23 pair is of particular interest as \isot{Na}{23} is relatively abundant in the white dwarf at the onset of carbon burning, $\Xisotm{Na}{23} \approx 10^{-4}$ \citep{martinez-rodriguez2016, piersanti2017, schwab2017a}, and \isot{Na}{23} is a primary product of the central carbon burning. 
    The A=21 Urca shell is approximately located at a density $\rho_{\rm{21}} \sim 3.75 \times 10^9 \, \gcc$, the A=23 Urca shell is at $\rho_{\rm{23}} \sim 1.85 \times 10^9 \, \gcc$, and the A=25 Urca shell is at $\rho_{\rm{25}} \sim 1.31 \times 10^9 \, \gcc$ \citep{suzuki2016}. 

	Convective mixing continually drives the Urca process in the simmering white dwarf, but convection is also affected by the Urca reactions in a number of ways. 
    Most straightforward is the creation of compositional gradients that impede convective mixing and have been shown to create semi-convective regions (see PA2026).
    Additionally, the Urca reactions directly alter the electron fraction of the fluid by adding electrons in the outer regions of the star and removing electrons in the interior.
    This change in the electron fraction has two consequences. 
	First, there forms a gradient of electrons that tend to flow from the exterior inward to the center.
    This flow runs counter to the electron chemical gradient, which largely follows the density gradient.
    Secondly, in a related manner, the buoyancy of the fluid is altered when the Urca reactions change the electron fraction, as it is the electrons which provide the pressure support in a degenerate white dwarf.
	The net effect is less buoyant fluid inside the Urca shell and more buoyant fluid outside the Urca shell.
    The importance of the compositional gradients and the effects from changes in the electron fraction is not well understood, but the general trend is that the convective Urca process restricts the convective mixing, reducing the kinetic energy of the star.

    It is additionally important to note the energetics of the Urca reactions. 
    Because the convection zone efficiently mixes the Urca pair across the Urca shell, the distribution of the Urca pair is out of local equilibrium.
    At local equilibrium, the Urca pair would be perfectly split with only electron capture products inside the Urca shell and beta-decay products outside the Urca shell.
    Instead, we have a more uniform distribution of the Urca pair throughout the convection zone, assuming moderately quick mixing.
    A consequence of this distribution being out of equilibrium is the Urca reactions will be exothermic throughout the star, except very close to the Urca shell.
    Even with the leakage of the neutrinos from the star, a net energy is produced in both the electron capture regions near the center and the beta-decay regions near the convective boundary.
    This can be understood as the consequence of a non-local-equilibrium distribution that releases energy as the Urca reactions relax the distribution back to local equilibrium.
	In particular, the difference in the threshold energy of the Urca nuclei and the electron chemical potential is deposited into the medium as excess thermal energy.
    This excess energy is readily supplied by the convection zone, which must do work to maintain the uniform non-local-equilibrium distribution.
    As in relation to the discussion above, the process in which this work is done is complex and should relate to the changing of the electron fraction.

    Studies of 2D simulations \citep{stein-wheeler2006} and analytic approaches \citep{lesaffre2005} have demonstrated that the convective Urca process may slow the convective flow and effectively reduce the total kinetic energy in the white dwarf. 
    To test and properly characterize these theories and effects, a 3D hydrodynamic model is needed to properly model the turbulent convection linked with the Urca reactions.
	Previous implementation of a 3D model in B2025 and PA2026 indicate less convective hindering than the previous work, but PA2026 did find the size of the convection zone to largely be reduced by about 30\%.
	The key finding in the two 3D models is the convection zone was not constrained tightly to the A=23 Urca shell as predicted by previous work.
	
	A key question of the convective Urca process is the degree, or extent, to which convection is impeded or slowed down. 
	A secondary question is how can we best characterize this effect and properly track the flow of energy?
    Ideally, we would be able to construct a model which would work with 1D stellar evolution codes, which can track the full simmering timescale.

\section{Numerical Methodology}\label{sec:sim_details}
    We use version 24.08 of \maestro\ \citep{fan2019}, a massively parallel low Mach number hydrodynamic code built on the \amrex\ framework \citep{zhang2019}, to simulate the convective core. 
    \maestro\ was designed to study stellar interiors and atmospheres, which are often low Mach number environments (i.e.\ the sound speed is fast compared to the characteristic fluid velocity).
    Modeling low Mach number flow with a more standard compressible finite-volume code is numerically challenging because the timestepping is limited by the acoustic wave timescale as opposed to the characteristic advective timescale.
    This limit on the timestep makes modeling slow moving fluids inefficient and often impossible for very low Mach numbers, the presented simulations have a Mach number of ${\sim} 0.005$.

    \maestro\ solves the following low Mach number hydrodynamic equations, paired with a nuclear reaction network, which correspond to the evolution of mass, momentum and enthalpy under a divergence constraint:
	\begin{gather}
	    \label{eqn:mass}
	    \frac{\partial (\rho X_k) }{\partial t} = {-}\nabla \cdot (\rho \mathbf{U} X_k) + \rho \dot{\omega}_k  \\
	    \label{eqn:momentum}
	    \frac{\partial \mathbf{U}}{\partial t}  = - \mathbf{U} \cdot \nabla \mathbf{U} - \frac{\beta_0}{\rho} \nabla \bigg(\frac{p-p_0}{\beta_0}\bigg) - \frac{\rho-\rho_0}{\rho} g \mathbf{e}_r \\
	    \label{eqn:energy}
	    \frac{\partial (\rho h)}{\partial t} = - \nabla (\rho h \mathbf{U}) + \frac{D\rho_0}{Dt} + \rho H_{\rm{nuc}} \\
	    \label{eqn:div_constraint}
	    \nabla \cdot (\beta_0 \mathbf{U}) = \beta_0 \bigg(S - \frac{1}{\bar{\Gamma}_1 p_0}\frac{\partial p_0}{\partial t}\bigg) \\
	    \label{eqn:HSE}
	    \nabla p_0 = -\rho_0 g\mathbf{e}_r.
	\end{gather}
    Here, $\rho, \mathbf{U}, h$ are mass density, velocity and enthalpy respectively. 
    $X_k$ is the mass fraction of the $k$th isotope and $\dot{\omega}_k$ is the creation/destruction rate of that isotope. 
    Note that $X_k$'s are defined such that $\sum_k X_k = 1$.
    $\rho_0$ and $p_0$ correspond to the base state, essentially the angle-averaged background state of the star, which remains in hydrostatic equilibrium (HSE). 
    $\beta_0$ is a density-like variable that captures the background stratification.
	$g$ is the magnitude of the gravitational acceleration.
	$H_\mathrm{nuc}$ is the specific energy generation rate due to nuclear reactions.
	Lastly, $S$ is the source term to the divergence constraint that accounts for perturbations related to compositional changes and heating from reactions, and $\bar{\Gamma}_1$ is the lateral average of the first adiabatic exponent ($d \log p/ d \log \rho |_s$).
    This formulation enforces conservation of total energy at low Mach number in the absence of external heating or viscous terms  \citep{klein2012, vasil2013}.

    We use the ``full-star" geometry in \maestro, which places the star at the center of a large 3D cartesian grid.
	Note, for this geometry, the enthalpy equation (Eqn.\ \ref{eqn:energy}) is decoupled from the solution.
	Instead, we use the base state pressure, $p_0$, in conjunction with the EOS to define the temperature.
	This is done to keep the base state in both HSE and thermodynamic equilibrium.
	The heating effect from the nuclear reactions is still incorporated via the source term, $S$, in the divergence constrain (Eqn.~\ref{eqn:div_constraint}).
	For the case of plane-parallel geometry in \maestro, there is an adjustment to the divergence constraint that allows for the enthalpy equation to be coupled to the solution~\citep{malone2011}.

    Enabled through \amrex, we implement three layers of mesh refinement such that our computational resources are focused on modeling the convective core.
    This choice allows for a higher ``effective resolution" of $5 \, \mathrm{km}$ for the full convection zone.
	Although \maestro\ has the capability of adaptive mesh refinement, we do not change our refinement levels in the presented simulations.
    For further details on \maestro\ and the low Mach number algorithm see \cite{fan2019} and references within.
    
\begin{figure*}
        \centering
        \includegraphics[scale=0.75]{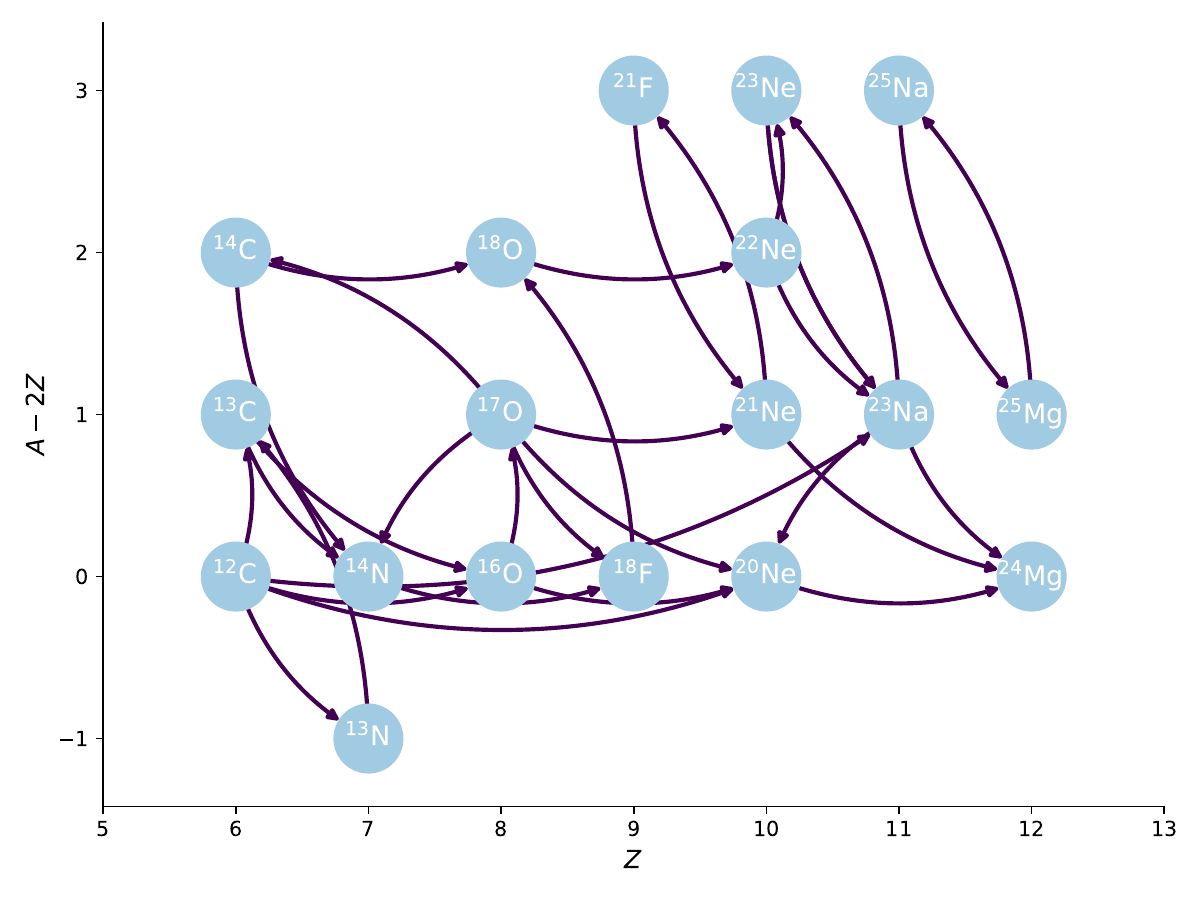}
        \caption{\label{fig:urca_net} 
            A graphical visualization of the nuclear reaction network. 
            Each node represents a different isotope in the network.
            Arrows represent reactions connecting two isotopes, with the direction of the arrow indicating the direction of the reaction.
            The horizontal axis indicates the proton number, Z. 
            The vertical axis indicates the number of excess neutrons (positive) or protons (negative). 
            Helium, protons, and neutrons are included in the network, but are excluded from this plot for the sake of clarity.
        }
    \end{figure*}

    A critical piece to modeling the convective Urca process is the coupling of nuclear reactions to the fluid motions.
    In \maestro\ the nuclear reactions are coupled to the fluid equations via Strang-splitting.
    The nuclear network implemented in this work is shown in Figure~\ref{fig:urca_net} and described further in Section~\ref{sec:network}.
	However, we note some of the additional microphysics here, which are implemented in C++ as part of the \microphysics\ project \citep{microphysics2024} that supports \amrex\ based simulation codes.
    We incorporate the effects of Coulomb screening on the reaction rates following \cite{graboske1973} for the weak limit and~\cite{alastuey1978, itoh1979} for the strong limit.
    We additionally include thermal neutrinos losses from the hot plasma following \cite{itoh1996}.
    We use a publicly available general purpose stellar EOS described in \cite{timmes2000, fryxell2000}, which takes into account the contributions of ions, electrons and radiation.

    In the low density outer regions, well outside the convection, we use a velocity sponge to dampen gravity wave-like excitations caused by the core convection, as described in \citep{zingale2009}.
    The sponge dampens the velocities toward zero by dividing by a constant factor, in our case $\kappa = 20$, for densities less than $\rho_{\mathrm{sponge}} = 3 \times 10^7 \, \gcc$.
    The sponge ensures that these low density regions, far outside the convection zone, do not limit our timestep, while not affecting the convection zone, where densities are of order $10^9 \, \gcc$.
    
\section{Reaction Network}\label{sec:network}
    We constructed a network to accurately track the carbon burning, both in compositional changes and energy generated. 
	To both explore various reaction networks and to implement this network with \maestro, we used the \pynucastro\ Python package~\citep{smith2023}.
    We started this process by creating a large network based on the network used in~\cite{piersanti2022}, with the exclusion of the A=56 isotopes, to use as a reference. 
    Using this network as a benchmark, we removed rates that did not meaningfully contribute to the energy generation rate or to compositional shifts over our ${\sim} 1 \, \mathrm{hr}$ timescale. 
    In order to test the network under simulation-like conditions, we created a unit test to mimic a fluid parcel cyclically traveling from the center of the star to the outer edge of the convection zone and back. 
    This allowed us to analyze which rates were important to the burning throughout the convection zone. 
	That is, the interior is dominated by the carbon burning, while further out, at lower temperatures and densities, the Urca reactions are far more significant.
    Our aim was to reduce the network to a size that we could compute using \maestro\ without significantly changing the energy generation rate or compositional shifts, particularly the neutronization due to burning. 
    We reduced our network from 120 rates down to 33 rates and 21 nuclei (see Figure~\ref{fig:urca_net}).

	The majority of rates in our network are strong rates related to the carbon burning and are taken from JINA  \texttt{REACLIB}~\citep{cyburt2010}.
	The weak rates related to the Urca reactions are tabulated rates from \cite{suzuki2016}.
    Note, our finished network is far larger than that used in previous work (B2025, PA2026), and better reflects the carbon burning at this stage of simmering.
    Particularly notable inclusions to the network are the addition of two more Urca pairs, \isot{Ne}{21} -- \isot{F}{21} and \isot{Mg}{25} -- \isot{Na}{25}, and the addition of more proton, neutron, and alpha captures onto \isot{C}{12}, \isot{O}{16}, and higher mass isotopes, which eliminates the need for any explicit $p \leftrightarrow n$ rates.
    Additionally, we include the $\isotm{N}{13}(e^{-},\nu_e)\isotm{C}{13}$ rate which is vital to properly capture the neutronization due to the carbon burning~\citep{chamulak2008, piro2008}.
	We use a tabulated version of this rate as in~\cite{piersanti2017, schwab2017a} computed using data from~\cite{zegers2008}. 
    
	Similar to PA2026, we ran two simulations with slightly different reaction networks.
	The first, Full Network simulation (FN) uses all the rates as displayed in Figure~\ref{fig:urca_net}.
	The second, No Beta simulation (NB) uses the same network with the exception of the three beta-decays in the Urca reactions.
	By removing the beta-decays from the network, we break the convective Urca process cycle and thus can better parse the effects of the convective Urca process on the FN simulation.
	We chose to keep the electron capture rates primarily because the A=23 electron capture is directly linked to the carbon burning in the center via $\isotm{C}{12}(\isotm{C}{12},p)\isotm{Na}{23}(e^{-},\nu_e)\isotm{Ne}{23}$. 
	The A=21 and A=25 electron capture rates are not similarly tied to the carbon burning but we include them for consistency across Urca pairs.

    Implementing this network in \maestro\ is computationally expensive, however the use of GPU acceleration has helped offset the cost of computing a larger network.
    One of the main pressure points is now the large number of unique isotopes, which slows the hydrodynamic calculations. 
    We look to address this further by using features implemented in \pynucastro\ to construct approximate networks which can eliminate intermediate nuclei, simplifying our network.
    
    This network was specifically designed for the following simulations, in which we only track ${\sim}1$ hour of the simmering phase.
	Studies of the longer timescale of the full simmering phase likely need a larger network to accurately track the stellar evolution and compositional changes.
	For higher temperatures, we expect this network to deviate more from our benchmark. 
	Similarly, we expect that future work to investigate lower central temperatures (i.e.\ earlier in simmering) will be able to reduce the number of isotopes and rates without losing accuracy in the carbon burning. 

\section{Initial Model}\label{sec:init_models}
     
    \begin{figure}
        \centering
        \includegraphics[width=\linewidth]{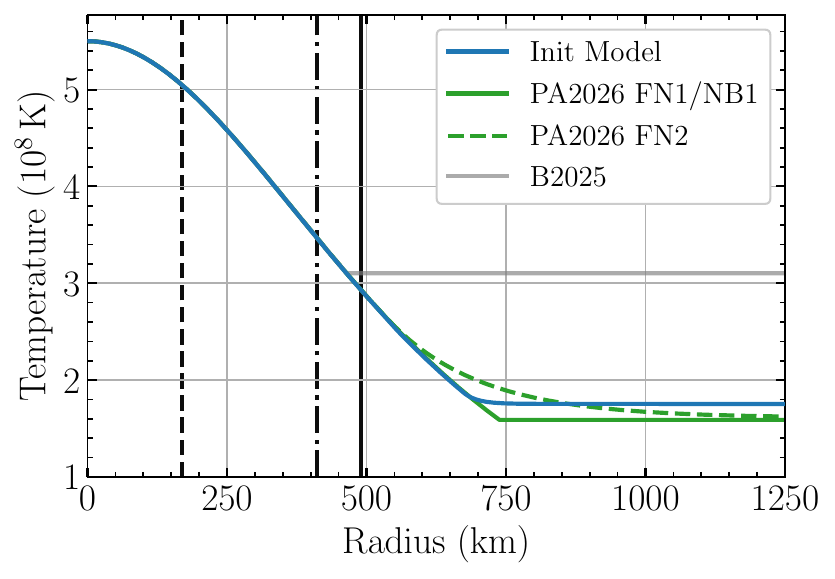}
        \caption{\label{fig:init_Tprofs} 
            Temperature profiles vs.\ radius for a series of initial models. 
            The blue curve represents the initial profile used in the presented simulations.
            The green curves represents the initial profile used in PA2026, solid representing the FN1/NB1 model and dashed representing the FN2 model.
            The grey curve represents the initial profile used in B2025.
            The vertical black lines indicate the location of the A=21 (dashed), A=23 (dot-dashed), and A=25 (solid) Urca shells.
        }
    \end{figure}

    \begin{figure}
        \centering
        \includegraphics[width=\linewidth]{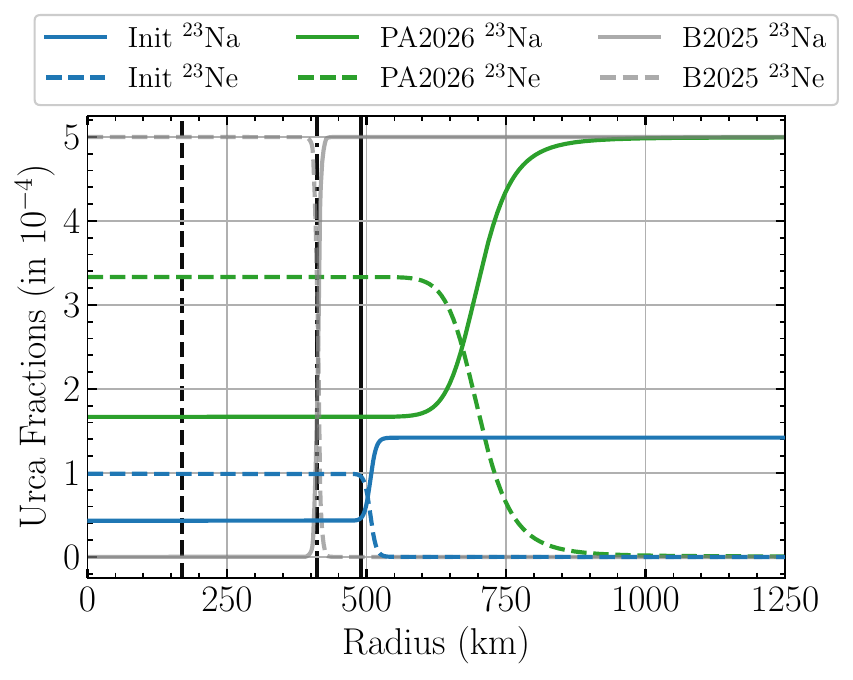}
        \caption{\label{fig:init_urca} 
            Mass fraction of the A=23 Urca pair vs.\ radius for a series of initial models. 
            Solid curves represent \isot{Na}{23} and dashed curves represent \isot{Ne}{23}.
            The blue curve represents the initial profile used in the presented simulations.
            The green curves represents the initial profile used in PA2026, only the FN1/NB1 initial model is shown.
            The grey curve represents the initial profile used in B2025.
            The vertical black lines indicate the location of the A=21 (dashed), A=23 (dot-dashed), and A=25 (solid) Urca shells.
        }
    \end{figure}

    To construct our parameterized initial model, we start with a central density and temperatures ($\rho_\mathrm{c} = 4.5 \times 10^9 \, \gcc$ and $T_{\mathrm{c}}=5.5 \times 10^8 \, \mathrm{K}$).
    This uses the same central temperature and density as previous work (B2025; PA2026) for easier comparison.
    From these central conditions, we integrate radially outward, maintaining HSE and following a specific temperature gradient. 
    For the interior region, we follow an adiabatic temperature gradient to mimic an established convection zone. 
    At a specified mass coordinate, we then switch to an isothermal temperature gradient that is stable to convection.
    An additional feature to the presented initial models is that of a smooth transition from the steep adiabatic gradient to the constant isothermal gradient. 
    This smoothing is done by limiting the rate at which the temperature gradient can change by a set factor.
	In our initial model presented here this factor is about 5\%, which creates a ${\sim}20 \, \mathrm{km}$ region that connects the main isentropic and isothermal regions.

    In Figure~\ref{fig:init_Tprofs}, we show the temperature profiles used in this paper as well as the initial temperature profiles used in previous work (B2025; PA2026).
    The dashed green curve and solid blue curve have a smoothed transition while the solid green curve and solid grey curve have no smoothing and instead sharply transition from adiabatic to isothermal.
    Note the temperature profile we implement in these simulation, solid blue curve, has a much smaller smoothed region compared to the PA2026 FN2 model.
	This relatively small smoothed region better reflects the expected profile for a simmering white dwarf as opposed to a large region that spans hundreds of kilometers.
    The primary goal of including this smooth transition is to reduce any discontinuities in the early stages of the simulation. 
    Smoothing out any discontinuities helps prevent excitations from arising and dominating the outer, convectively stable regions of the white dwarf.
    \maestro\ is highly sensitive to any discontinuities in the initial conditions which can make the initial evolution numerically challenging.

	In our initial model, we set the isentropic region to encompass about $0.9 \, \Msun$ which sets the isothermic temperature at $T_\mathrm{iso} \approx 1.8 \times 10^8 \, \mathrm{K}$.
	This $T_\mathrm{iso}$ is consistent with 1D stellar evolution models, which typically range around $10^8 \, \mathrm{K}$, depending on the accretion and cooling history of the progenitor model \citep{martinez-rodriguez2016, piersanti2022}.
	The size of this isentropic region compares fairly well with the PA2026 FN1/NB1 initial model, which has a isentropic zone of $1 \, \Msun$.
	Importantly, the initial isentropic region extends past all three Urca shells present in our simulation.

    In addition to setting the temperature and density profiles, we also specify the compositional distribution. 
    We distribute the Urca nuclei similarly to PA2026, attempting to reflect the dynamic equilibrium distribution of the Urca pairs.
    This is done by distributing the Urca nuclei following a quick mixing limit approximation~\citep{boydproceedings}.
	The distribution for the A=23 Urca pair is shown in Figure~\ref{fig:init_urca}, along with the initial distributions from previous work.
	One difference from PA2026 is where we locate the transition of the Urca pair.
	In these new models, we move the transition point about $200 \, \mathrm{km}$ inward from the initial isentropic boundary.
    Placing the Urca transition too close to the initial convective boundary can result in a small portion of the beta-decay reactant (e.g.\ \isot{Ne}{23}) distributed into the isothermal region.
    This material quickly beta-decays and produces excess energy at the beginning stages of the simulation and can result in unwanted behavior in the convectively stable regions.
    To prevent this, we place the transition in the Urca nuclei further inward to ensure the mass fraction of the beta-decay reactant is zero at the isentropic boundary. 
	Note the total mass fraction for a given Urca pair is constant throughout the white dwarf (e.g.\ $\Xisotm{Ne}{23} + \Xisotm{Na}{23} = 1.42 \times 10^{-4}$ everywhere). 

    A new feature that we add, which differs from previous work, is a small \isot{C}{12}-\isot{O}{16} gradient.
	This gradient is included in an attempt to better reflect a white dwarf that has existing CO gradients and had burned some of the carbon to this point.
	This gradient of \isot{C}{12}-\isot{O}{16} is placed at the same location as the Urca gradients ($R\approx 500 \mathrm{km}$), such that all the compositional gradients aligned in the initial model.

    The composition for the initial model, shown in Table~\ref{tab:init_comp}, are based on Table~1 in~\cite{schwab2017a}.
    All non-listed isotopes are initialized with zero mass fraction.
    The composition presented in \cite{schwab2017a} represents the composition near the start of the simmering phase, thus it is not fully accurate to this later stage of simmering.
	However, this still reflects a fairly reasonable initial composition for our simulations and improves upon previous work which assumed a fully CO white dwarf with the only addition of a small fraction of the Urca pair (B2025; PA2026).
	A relevant consequence of our choice in initial model is the relatively small abundance of the A=23 Urca pair.
	Since \isot{Na}{23} is produced in carbon burning, we would expect this Urca pair to be more significant at this stage of simmering.
	We will discuss the significance of this relatively small abundance in further detail as well as compare to previous work which included almost three times the amount of the A=23 Urca pair.

	In future work we hope to explore more complex initial models that will better reflect the prior evolution. 
	This include accurately factoring in changes to composition due to carbon burning, as well as the influence compositional gradients have on the temperature gradient.
	In particular, the potential for compositional gradients and the convective Urca process to hinder convection and produce large super-adiabatic convective boundaries, such as seen in \cite{piersanti2022}.

    \begin{table}
		\centering{
        \begin{tabular}{l c c}
            \hline \hline
            Isotope      & $X(R < 500 \, \mathrm{km})$ & $X(R > 500 \, \mathrm{km})$ \\
            \hline 
            \isot{C}{12} & 0.4071 & 0.04076     \\
            \isot{C}{13} & $4.0 \cdot 10^{-5}$ &  \\ 
            \isot{O}{16} & 0.5788 &  0.5784 \\
            \isot{F}{21} & $3.736 \cdot 10^{-8}$ & 0.0 \\ 
            \isot{Ne}{20} & $1.34 \cdot 10^{-4}$ &  \\ 
            \isot{Ne}{21} & $3.736 \cdot 10^{-5}$ & $3.740 \cdot 10^{-5}$ \\ 
            \isot{Ne}{22} & $1.37 \cdot 10^{-2}$ &  \\
            \isot{Ne}{23} & $9.467 \cdot 10^{-5}$ & 0.0 \\
            \isot{Na}{23} & $4.733 \cdot 10^{-5}$ & $1.42 \cdot 10^{-4}$ \\
            \isot{Na}{25} & $3.84 \cdot 10^{-5}$ & 0.0 \\
            \isot{Mg}{25} &  0.0 & $3.84 \cdot 10^{-5}$ \\
            \hline \hline
        \end{tabular}}
        \caption{\label{tab:init_comp} The initial model composition. 
            The isotopes that varied, \isot{C}{12}, \isot{O}{16} and Urca pairs, are defined twice for their asymptotic behavior.
            Those that remain constant throughout the domain are only listed once. All other isotopes are initialized with an abundance of zero.}
    \end{table}

\section{Results \& Analysis}\label{sec:analysis}
	In the following subsections, we discuss the results of our Full Network simulation and No Beta simulation.
	In subsection~\ref{subsec:init_evol}, we discuss the initial evolution with which we start the simulation without evolving the base state.
	Next, in subsection~\ref{subsec:steady_state}, we analyze how the FN and NB simulations relax to a steady state.
	In subsection~\ref{subsec:conv_struct}, we compare the convective structure of the FN and NB simulations.
	Then subsection~\ref{subsec:urca_in_conv} we analyze the Urca process and its effects on the convection zone.
	Lastly, in subsection~\ref{subsec:comp_f} we compare the FN simulation to the results of PA2026.

    \subsection{Initial Evolution}\label{subsec:init_evol}

		In previous work, PA2026, we found small near-random velocities in the convectively stable regions of the simulations. 
		These developed early on, when the convection zone had not yet been established.
		Once the convection zone had been established, the interaction between the convection zone and the convectively stable boundary produced wave-like oscillations, which largely replaced these random velocities. 
		However, these oscillating velocities still dominated some low-resolution regions of the simulation.
		We did not find this phenomena present in B2025, likely because these simulations did not evolve the base state.

		In an attempt to prevent the early development of these random velocities, we ran an initial evolution of the convection zone with the base state held constant.
		For this initial evolution, we use the full reaction network, including the beta-decays, and stopped the simulation when the convection zone appeared to largely be established, about 885 seconds.
        The resulting convective velocity field was dominated by a large velocity dipole, and the composition in the convection zone was fairly well mixed.

\begin{figure}[htb]
            \centering
            \includegraphics[width=\linewidth]{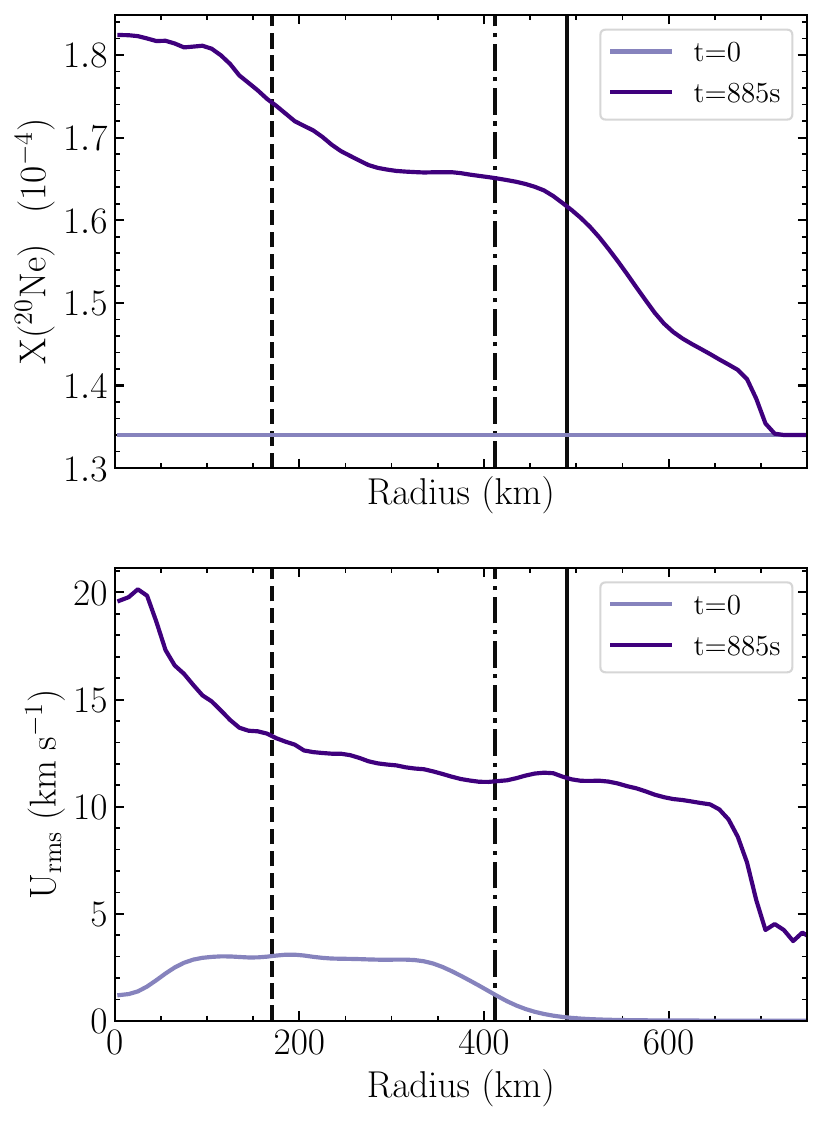}
            \caption{\label{fig:pre_bstate} 
                The top plot shows the angle-averaged \Xisot{Ne}{20} vs.\ radial bin.
                The bottom plot shows the rms velocity vs.\ radial bin.
                The light purple indicates the t=0 initial conditions.
                The dark purple curve represents the simulation at the end of the initial evolution, about $t=885 \, s$.
            	The vertical black lines indicate the location of the A=21 (dashed), A=23 (dot-dashed), and A=25 (solid) Urca shells.
            }
        \end{figure}

        The \isot{Ne}{20} and \vrmsr\ profiles are displayed in Figure~\ref{fig:pre_bstate}.
        We use \isot{Ne}{20} as a tracer of compositional mixing because it is a main product of the carbon burning and it does not significantly interact with any of the Urca pairs.
        The composition profile shows a relatively well mixed convection zone, with exceptions near the center, where carbon burning is producing \isot{Ne}{20}, and near the convection boundary, where two distinct downward slopes have formed.
        In the velocity profile, we find a relatively constant rms velocity throughout the zone that then drops near the convective boundary.
        The composition and velocity profiles align reasonably well with the isentropic boundary of the initial model. 
        However, the specifics of each profile should not be overstated as the simulation has not relaxed to a steady state and the base state has been held constant.
        We present these profiles here to indicate the nature of the convection zone which works as the initial state for the FN and NB simulations. 

        Overall, after evolving to 885 seconds, we found a convective velocity structure similar to that found in~\cite{boyd2025}, which also did not evolve the base state.
        Without the base state evolution, we do not track any of the expansion of the star due to the carbon burning heating the core.
		This expansion is not appreciably large, but it is significant enough to change the dynamics of the convection zone.
        Because the base state is held constant, the difference between the density, $\rho$, and base state density, $\rho_0$, steadily increases.
        Note that $\rho$ does track some of the expansion even while $\rho_0$ remains constant.
        This artificially boosts the buoyancy term in Equation~\ref{eqn:momentum} as the difference $\rho - \rho_0$ increases in magnitude.
        As we see in Section~\ref{subsec:steady_state}, properly tracking the base state will lead to a shift in the velocity structure and in particular the magnitude of the velocities in the convection zone.

        By starting without the base state, we were able to reduce some of the near-random velocities in the stable isothermal envelope surrounding the convection zone.
        We still found that some of these velocities developed once the base state evolution was turned on, however this only affected low resolution regions well away from the convection zone.
        Because these velocities only developed in lower resolution regions, it is possible that we are not fully resolving some excitations in the convectively stable regions, which are driving these velocities.

		An additional benefit of this ``initial evolution" phase was the saving of some computational resources.
		The first thousand seconds of these convective simulations mainly deal with establishing an initial convective velocity field.
        By only evolving a single simulation over this period, we save computational resources that can instead be spent accurately modeling the simulations as they relax to a steady-state.
        A similar procedure to develop an initial convective velocity field has been shown to work well in massive star simulations~\citep{zingale2024}. 
		In future work, we look to explore these methods for setting the initial convective velocity field while saving computational resources.

    \subsection{Settling to steady state}\label{subsec:steady_state}
\begin{figure*}[htp]
            \centering
            \includegraphics[width=\linewidth]{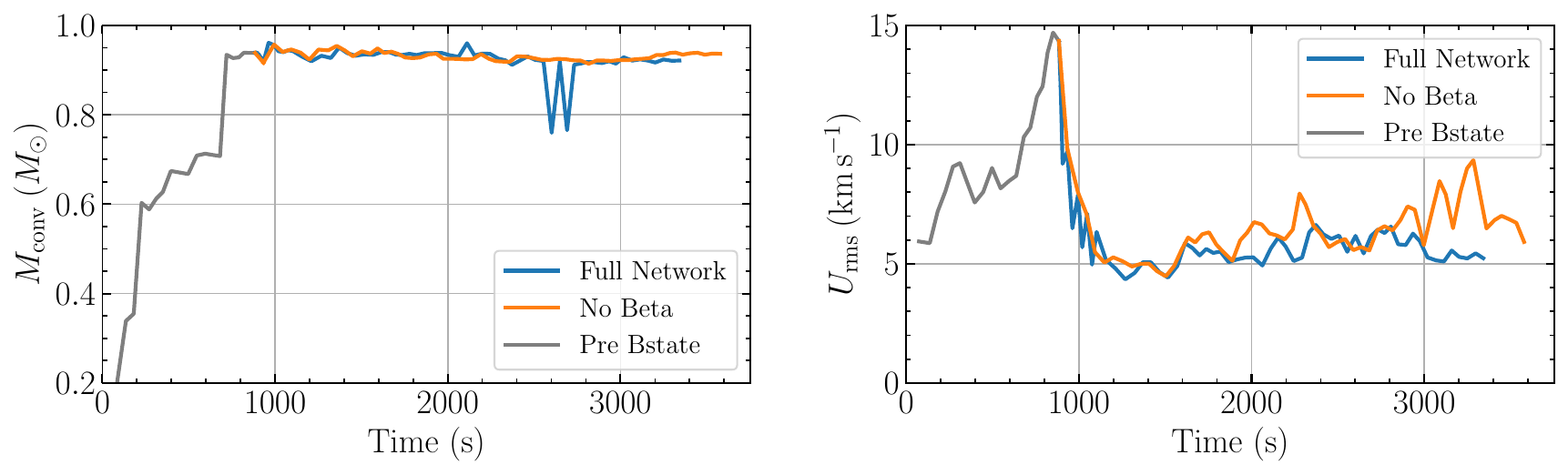}
            \caption{\label{fig:mconv_vrms} 
                The two plots tracks the mass enclosed by the convection zone (left) and rms velocity in the convection zone (right) over time.
                The grey curve shows the initial evolution, to $t=885 \, \mathrm{s}$, prior to evolving the base state.
                The blue curve represent the FN simulation and the orange curve represents the NB simulation.
            }
        \end{figure*}

        Using the results of the initial evolution described above as our initial state, we ran the FN and NB simulations for over $2500 \, \mathrm{s}$ corresponding to tens of convective turnovers.
		The velocity field and broader convective structure of the initial state is quickly replaced within a few convective turnover times.
        We stop each simulation around the same time as both had reached a steady state.
        We determine the simulations have settled to a steady-state when the size of the convection zone and the characteristic convective velocity settle to near constant values.
        This is shown in Figure~\ref{fig:mconv_vrms}, where the evolution of the mass interior to the convection zone, \Mconv, and the rms velocity, \vrms, are shown over time.
        We define the extent of the convection zone using compositional gradients, similar to previous work (B2025; PA2026).
		We first calculated a 1D profile using mass weighted angle-averages of \Xisot{Ne}{20}, and then we calculated the radial gradient. 
		The convective boundary is defined to be at the minimum of this radial gradient.
		The only change from previous study is the use of \isot{Ne}{20} as opposed to \isot{C}{12}.
		We made this change because \isot{Ne}{20} is uniformly distributed in our initial model, unlike \isot{C}{12}.
		The uniform distribution makes tracking the extent of compositional mixing more straightforward, particularly at earlier times in the simulation when the initial distribution is still important.

        The rms velocity is defined for the convection zone as,
        \begin{equation}
            \vrms = \sqrt{ \langle {U_x} ^2 \rangle + \langle {U_y} ^2 \rangle + \langle {U_z} ^2 \rangle } ,
        \end{equation}
        where $U_x, U_y, U_z$ are velocity components in the $x,y,z$ direction. 
        The angled brackets $\langle \rangle$ represent a density-weighted average over the convection zone.
		With the size and velocity of the convection zone, we can estimate the convective turnover time as $\Tauconvm \approx 2 \Rconv / \vrms \approx 250 \, \mathrm{s}$.

        In the initial evolution phase, prior to evolving the base state, the convective core approximately settles to a size of $\Mconv \approx 0.95 \, \Msun$.
        The rms velocity also increases steadily and peaks at the end of this initial evolution but quickly drops off once we begin the FN and NB simulations, which evolve the base state.
        As mentioned in Section \ref{subsec:init_evol}, the \vrms\ decreases because we now evolve the base state and track the slight expansion of the star due to the added heating from central carbon burning.
        There is not a corresponding change in the convective mass because the mixing continues to extend into the outermost regions, just at a slower rate.
		Once we evolve the two simulations with the base state evolution, the rms velocity corresponds much more closely to the findings in PA2026.
		This result demonstrates the importance of evolving the base state to properly model the convection zone during this stage of simmering.
        
		In the FN simulation, there are two significant dips in the convective mass around 2600 seconds.
        These dips are due to a region temporarily having an even steeper compositional gradient.
		That a region like this can exist is indicative of the difference between the convective boundaries of the FN and NB simulations, which we will expand on later.
		Note that there is an inherent uncertainty or ambiguity in the radius of the convection zone. 
        The convective boundary has a width, particularly in the FN simulation which we explore more in Section~\ref{subsec:conv_struct}.

\begin{figure}
            \centering
            \includegraphics[width=\linewidth]{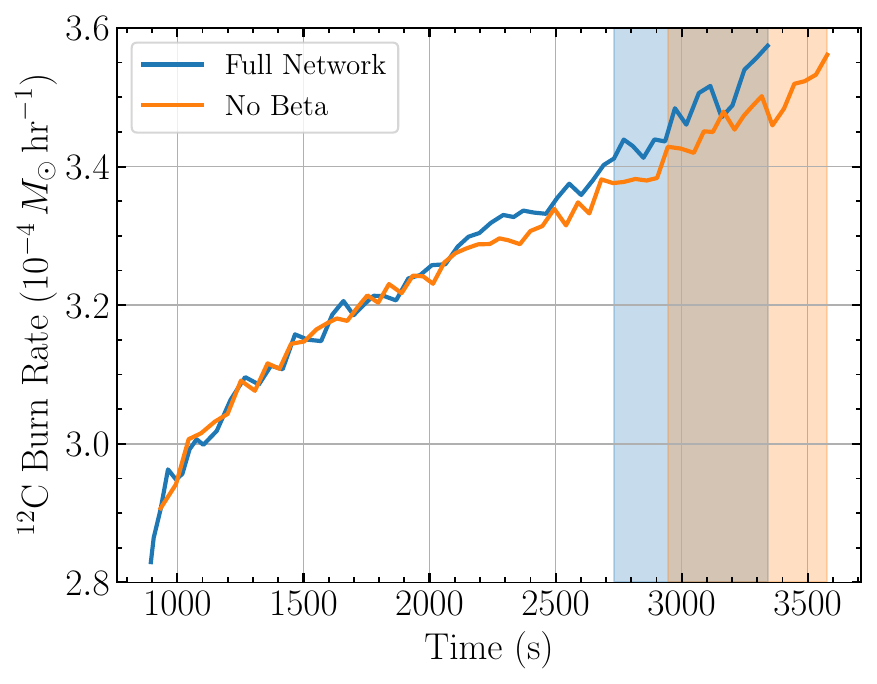}
            \caption{\label{fig:cburn_rate} 
                Rate of \isot{C}{12} mass burned in the simulations over time. 
                The blue curves represents the FN simulation while the orange curves represents the NB simulation. 
                The shaded regions correspond to the time intervals with which we calculate time-averages of some parameters.
            }
        \end{figure}

        One aspect that does not settle to a steady-state is the carbon burning rate (see Figure~\ref{fig:cburn_rate}).
        As we are studying the simmering phase, we expect the carbon burning rate to steadily increase over time.
        The white dwarf is fundamentally in a state of thermonuclear runaway, though it is still a very slow runaway at this point.
        In order to further compare the simulations, and to smooth out any temporary variations, we look at time averaged variables for each simulation in subsequent sections.
		The time spans are indicated by the shaded regions in Figure~\ref{fig:cburn_rate}, and account for the last ${\sim}600$ seconds of simulation time (2-3 convective turnovers).
        We ensure the comparison is consistent in terms of the rate of carbon burning; the average carbon burning rate during these two intervals differ by less than 0.01\%.

    \subsection{Convective Structure \& Boundary} \label{subsec:conv_struct}
\begin{figure*}
            \centering
            \includegraphics[scale=0.5]{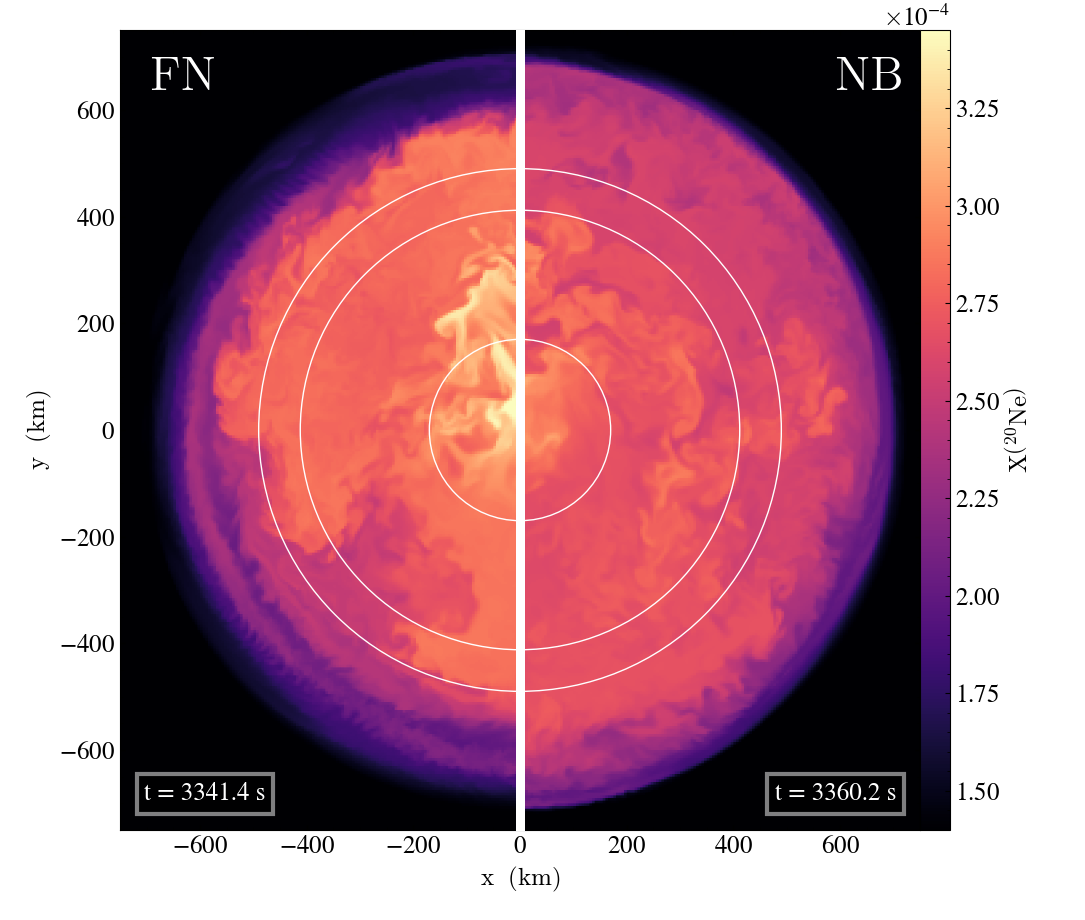}
            \caption{ \label{fig:ne20_slice} 
                Slices through the center of the white dwarf, zoomed in on the convection zone. 
                The colorbar represents the mass fraction of \isot{Ne}{20}.
                The left half of the slice shows the FN simulation and the right half shows the NB simulation. 
                The three white circles indicate the location of the three Urca shells, A=21, A=23, and A=25 (from innermost to outermost).
            }
        \end{figure*}

\begin{figure*}
            \centering
            \includegraphics[scale=0.5]{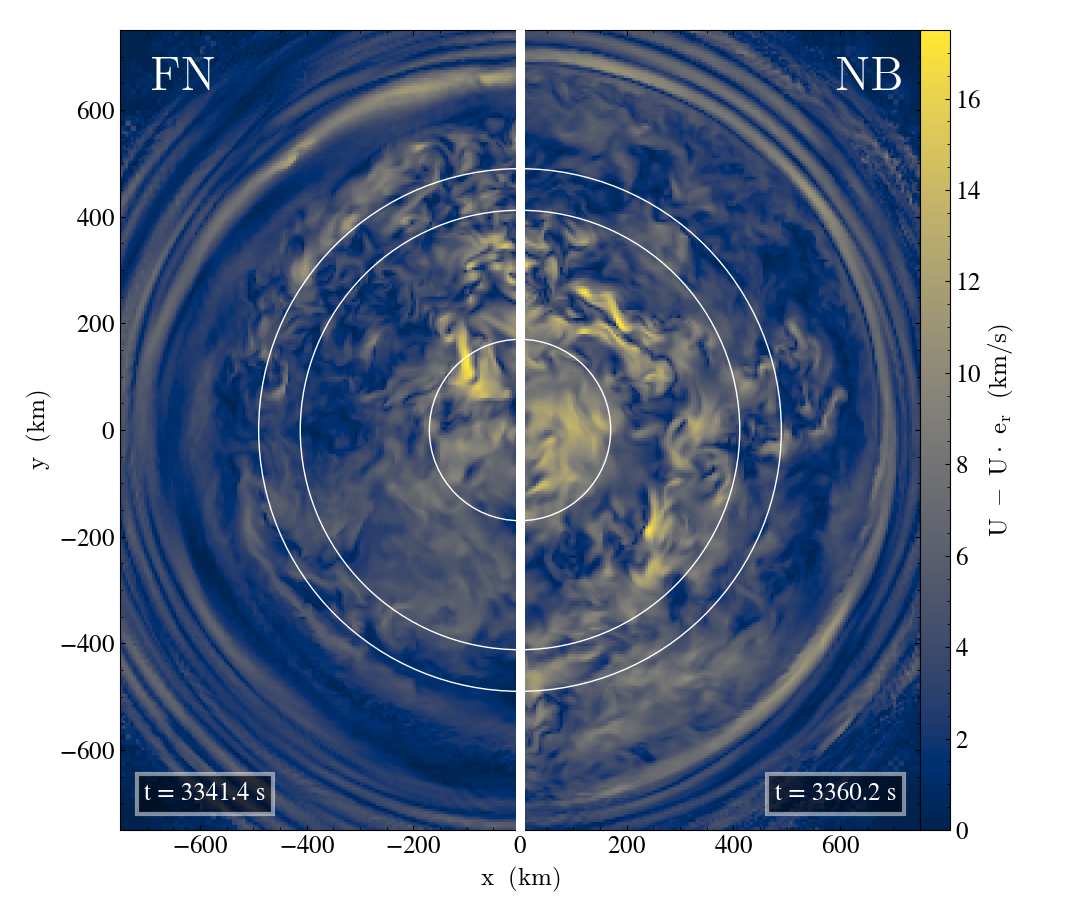}
            \caption{ \label{fig:tanvel_slice} 
                Slices through the center of the white dwarf, zoomed in on the convection zone. 
                The colorbar represents the magnitude of the tangential velocity.
                The left half of each slice shows the FN simulation and the right half shows the NB simulation. 
                The three white circles indicate the location of the three Urca shells, A=21, A=23, and A=25 (from innermost to outermost).
            }
        \end{figure*}

		The FN and NB simulations appear to settle to similar convection zone sizes based on their respective \Mconv\ and \vrms (see Figure~\ref{fig:mconv_vrms}).
		However, clear differences emerge when we look more closely at the convective structure.
		In Figure~\ref{fig:ne20_slice} and~\ref{fig:tanvel_slice}, we plot 2D slices of \Xisot{Ne}{20} and the tangential velocity for both the FN and NB simulations.
		There is a stark difference in how carbon burning products, like \isot{Ne}{20}, are mixed throughout the convection zone.
		The NB simulation indicates relatively uniform mixing out to about $700\, \mathrm{km}$, with a clear convective boundary.
		The FN simulation instead shows mixing that is roughly uniform out to about $550 \, \mathrm{km}$, with a more non-uniform region from 550-$700\, \mathrm{km}$.
		A similar picture can be seen in the tangential velocity, though the additional gravity wave-like features in stable regions and more chaotic nature of the velocity field makes the comparison far less clear.

\begin{figure}
            \centering
            \includegraphics[width=\linewidth]{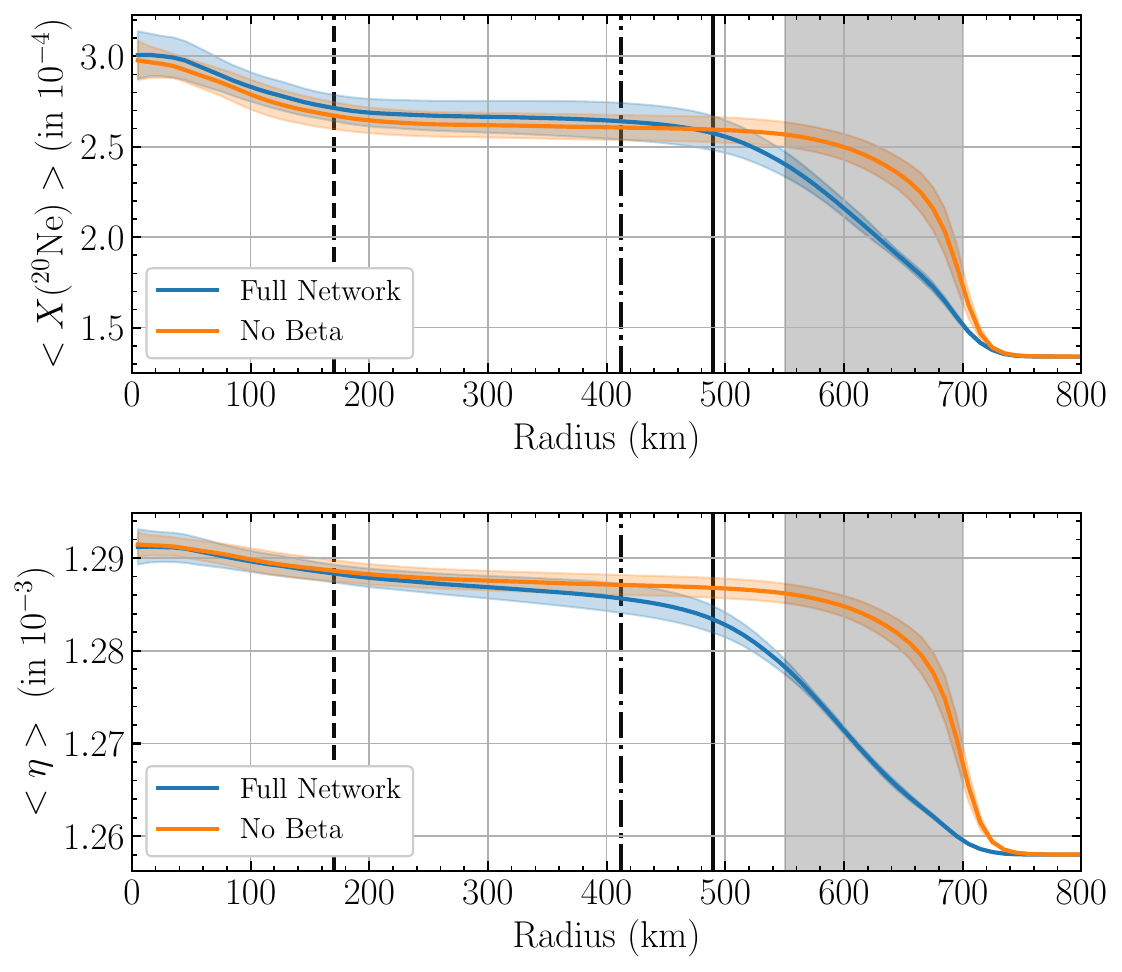}
            \caption{\label{fig:ne20_eta_profs} 
                Density-weighted angle-averaged profiles of \Xisot{Ne}{20} (top) and neutron excess, $\eta$, (bottom) vs.\ radial bin. 
				The blue curves represent the FN simulation, while the orange represent the NB simulation.
				Each curve represents the time averaged value for the given simulation over the last ${\sim}600 \, \mathrm{s}$, see Figure~\ref{fig:cburn_rate}.
                The blue and orange shaded regions bounding each curve represents a single standard deviation away from the mean value.
				The grey vertical shaded regions represents the inefficiently mixed region, from 550-$700\, \mathrm{km}$, see Section~\ref{subsec:conv_struct}.
            	The vertical black lines indicate the location of the A=21 (dashed), A=23 (dot-dashed), and A=25 (solid) Urca shells.
            }
        \end{figure}

		Looking at averaged profiles further demonstrates the difference in mixing (see Figure~\ref{fig:ne20_eta_profs}).
		The top plot, showing the \Xisot{Ne}{20} profile, demonstrates the two simulations significantly diverge in the outer region from $550 \, \mathrm{km}$ to the extent of compositional mixing near $700\, \mathrm{km}$, see the shaded region.
		Because the abundance of \isot{Ne}{20} is largely distinct from the Urca reactions, it is also important to consider the more general compositional changes.
		The abundance of \isot{Ne}{20} is a good indicator of the mixing of carbon burned material, however the composition is also altered by the Urca reactions.
		To capture the broader compositional changes, we plot the neutron excess, $\eta = \sum_i X_i(N_i - Z_i)/A_i$ (see the bottom plot of Figure~\ref{fig:ne20_eta_profs}).
		Here, we can see the composition changes inward to the shaded region, due to the Urca reactions.
		Along with the 2D slices, these profiles support that the region 550-$700\, \mathrm{km}$ is less efficiently mixed in the FN simulation.

\begin{figure}
            \centering
            \includegraphics[width=\linewidth]{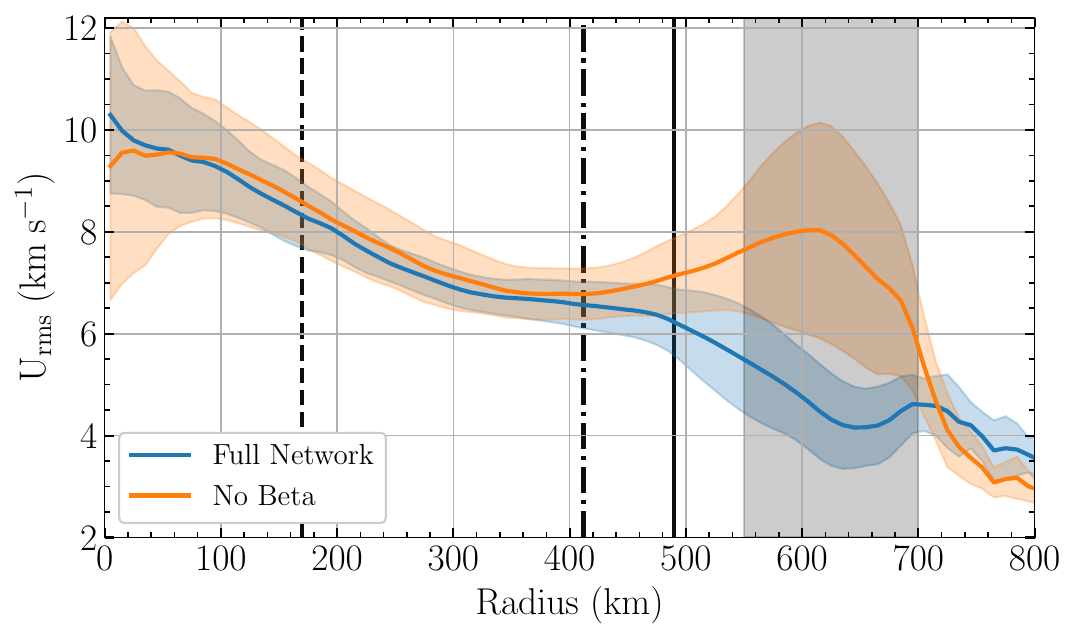}
            \caption{\label{fig:vrms_profs} 
                The plot shows the rms velocity vs.\ radial bin.
				The blue curves represent the FN simulation, while the orange represent the NB simulation.
				Each curve represents the time averaged value for the given simulation over the last ${\sim}600 \, \mathrm{s}$, see Figure~\ref{fig:cburn_rate}.
                The shaded regions bounding each curve represents a single standard deviation away from the mean value.
				The grey vertical shaded regions represents the inefficiently mixed region, from 550-$700\, \mathrm{km}$, see Section~\ref{subsec:conv_struct}.
            	The vertical black lines indicate the location of the A=21 (dashed), A=23 (dot-dashed), and A=25 (solid) Urca shells.
            }
        \end{figure}

		Looking at the velocity structure supports this idea of an inefficiently mixed outer region.
		We plot the \vrmsr\ vs.\ radius for each simulation in Figure~\ref{fig:vrms_profs}.
		Once again, the simulations are well aligned in the inner $500 \, \mathrm{km}$, but meaningfully diverge outside roughly $550 \, \mathrm{km}$.
		Note here, the turbulent convection is inherently chaotic and thus there is significant variance in our time averaged \vrmsr\ profiles.
		So though the \vrmsr\ profiles appear to diverge closer to $450 \, \mathrm{km}$, aligning more with the $\eta$ profiles, there is still significant overlap in the variance between the FN and NB simulations out to a radius of about $550 \, \mathrm{km}$.

		Lastly, it can be useful to compare our simulations to the Schwarzschild and Ledoux criteria for convective stability~\citep{HKT2004}.
		We follow the calculation detailed in \cite{boyd2025}, but summarize here.
		The Schwarzschild criterion compares the temperature gradient of the star to an adiabatic temperature gradient.
		This is formulated as
		\begin{equation}\label{eqn:schwarz}
		    \frac{d \log T}{d \log P} > \left( \frac{d \log T}{d \log P} \right)_\mathrm{ad} ,
		\end{equation}
		where it is common to denote the gradients as $\nabla$ and $\nabla_{\mathrm{ad}}$, respectively.

		When compositional gradients are present, they tend to stabilize regions against convection. 
		The Ledoux criterion compensates for this effect by adding an additional term to the comparison.
		This is formulated as
		\begin{equation}\label{eqn:led}
		    \frac{d \log T}{d \log P} > \left( \frac{d \log T}{d \log P} \right)_\mathrm{ad} + B ,
		\end{equation}
		where $B$ is a term that represent the compositional gradients, and can be defined with the mean molecular weight $\mu$, $B = (\chi_T/\chi_{\mu}) (d\log \mu / d \log P)$.
		Here, $\chi_{\mu} = \left( \partial \log P / \partial \log \mu \right)_{\rho, T}$ and  $\chi_{T} = \left( \partial \log P /\partial \log T \right)_{\rho, \mu}$ are thermodynamic derivatives.
		The right hand side of this inequality is commonly denoted as $\nabla_{\mathrm{Led}}$.
		We construct and calculate these gradients,$\nabla$, $\nabla_{\mathrm{ad}}$, and $\nabla_{\mathrm{Led}}$, using centered differences as described in~\cite{boyd2025}.

\begin{figure}
            \centering
            \includegraphics[width=\linewidth]{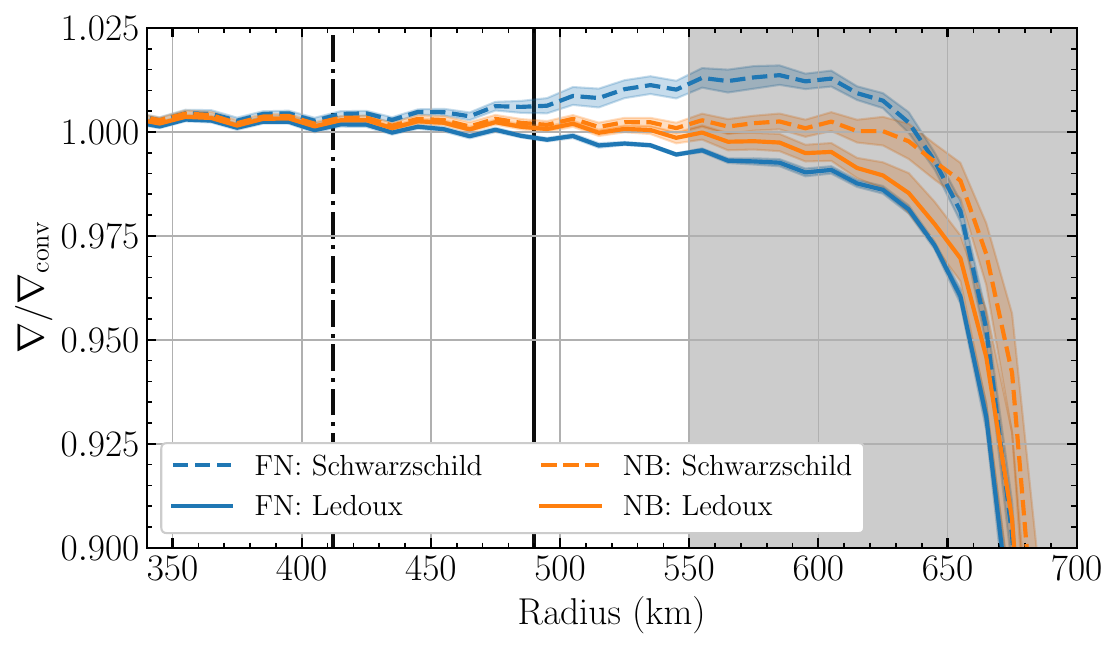}
            \caption{\label{fig:conv_grad} 
                Density-weighted angle-averaged profiles of the ratio of convective gradients vs.\ binned radius. 
				The dashed curves relate to the Schwarzschild criterion $\nabla / \nabla_{\mathrm{ad}}$.
				The solid curves relate to the Ledoux criterion $\nabla / \nabla_{\mathrm{Led}}$.
				The blue curves represent the FN simulation, while the orange represent the NB simulation.
				Each curve represents the time averaged value for the given simulation over the last ${\sim}600 \, \mathrm{s}$, see Figure~\ref{fig:cburn_rate}.
                The shaded regions bounding each curve represents a single standard deviation away from the mean value.
				The grey vertical shaded regions represents the inefficiently mixed region, from 550-$700\, \mathrm{km}$, see Section~\ref{subsec:conv_struct}.
            	The vertical black lines indicate the location of the A=23 (dot-dashed) and A=25 (solid) Urca shells.
            }
        \end{figure}

		In Figure~\ref{fig:conv_grad}, we plot the ratio of $\nabla$ to $\nabla_{\mathrm{ad}}$ and to $\nabla_{\mathrm{Led}}$.
		Comparing the two criteria for the FN simulation, we see a divergence outside the A=23 Urca shell.
		This could be considered a semi-convection zone as the region is unstable according to the Schwarzschild criterion but is stable according to the Ledoux criterion.
		The NB simulation has a somewhat similar structure closer to the convective boundary, where the build up of compositional gradients provide some stability to convection.
		The NB simulation, however, has no regions showing a super-adiabatic temperature gradient as seen in the FN simulation from 450-$640 \, \mathrm{km}$, see dashed blue curve.
		This indicates the compositional gradients produced by the Urca reactions, mainly the A=23 pair, are providing a stabilizing force that is altering the outer convective region.
  
\begin{figure}
            \centering
            \includegraphics[width=\linewidth]{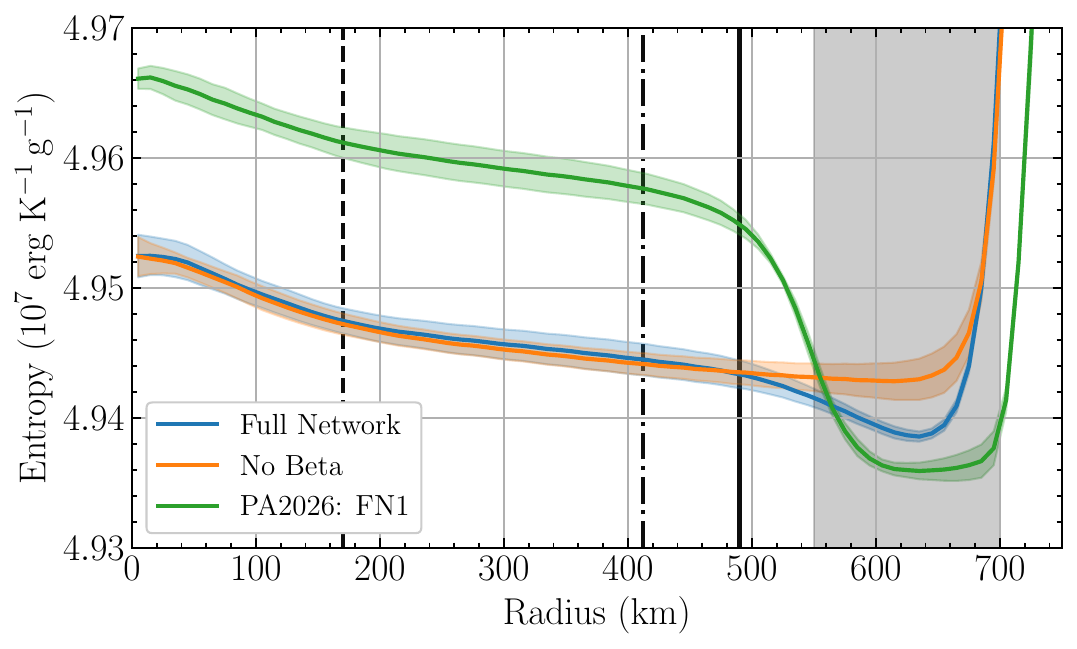}
            \caption{\label{fig:entropy} 
                Density-weighted angle-averaged profiles of the specific entropy vs.\ binned radius. 
				The blue curves represent the FN simulation, while the orange represent the NB simulation.
				Each curve represents the time averaged value for the given simulation over the last ${\sim}600 \, \mathrm{s}$, see Figure~\ref{fig:cburn_rate}.
                The green curve is from the PA2026 FN1 simulation with a similar time averaging over the last ${\sim}600 \, \mathrm{s}$ of the simulation.
                The shaded regions bounding each curve represents a single standard deviation away from the mean value.
				The grey vertical shaded regions represents the inefficiently mixed region, from 550-$700\, \mathrm{km}$, see Section~\ref{subsec:conv_struct}.
            	The vertical black lines indicate the location of the A=21 (dashed), A=23 (dot-dashed), and A=25 (solid) Urca shells.
            }
        \end{figure}

		A further demonstration of the effects of the convective Urca process on the convection zone is the changes to the entropy.
		The interior convection zone shows a reasonably flat entropy profile as one would expect for an efficiently mixed region.
		However, as noted in~\cite{sbw1999}, because the convective Urca process restricts convection, there  should be a dip in the entropy profile coinciding with the convective boundary.
		We see such a feature in our FN simulation as shown in Figure~\ref{fig:entropy}.
		Here, the dip in entropy aligns reasonably well with the inefficiently mixed region, marked by the shading.

	\subsection{Urca Process in the Convection Zone}\label{subsec:urca_in_conv}
\begin{figure*}
            \centering
            \includegraphics[width=\textwidth]{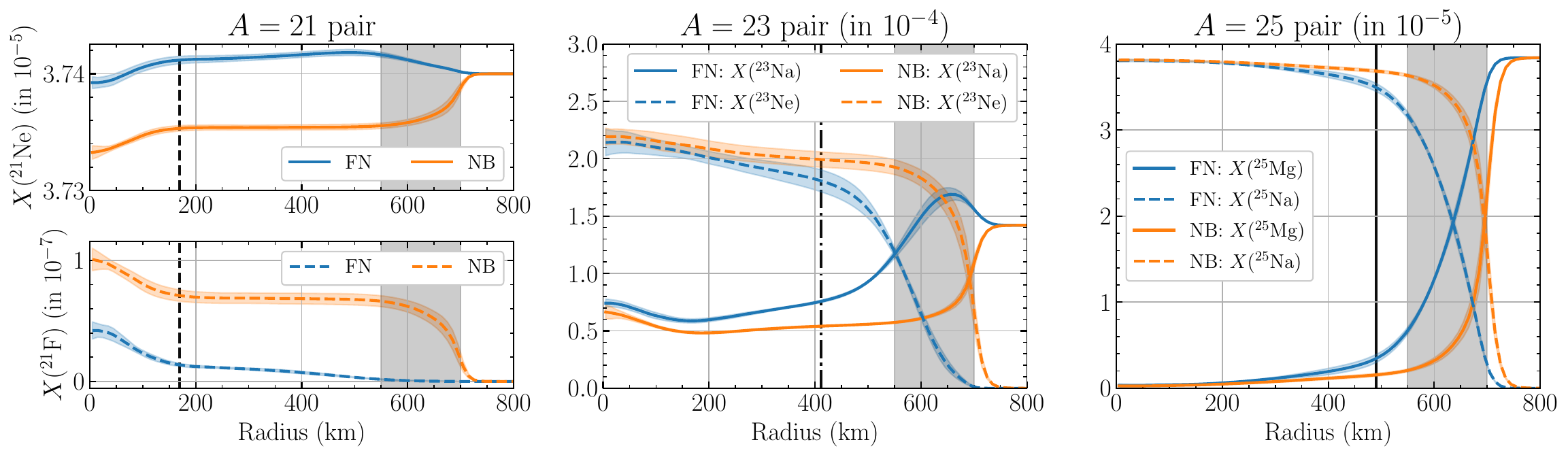}
            \caption{\label{fig:urca_comp_profs} 
                Density-weighted angle-averaged composition profiles for the A=21 (left), A=23 (center), and A=25 (right) Urca pairs vs.\ binned radius. 
            	Solid curves represent the relatively neutron-poor isotope (product of beta-decay) and the dashed curves represent the relatively neutron-rich isotope (product of electron capture).
				The blue curves represent the FN simulation, while the orange represent the NB simulation.
				Each curve represents the time averaged value for the given simulation over the last ${\sim}600 \, \mathrm{s}$, see Figure~\ref{fig:cburn_rate}.
                The blue and orange shaded regions bounding each curve represents a single standard deviation away from the mean value.
				The grey vertical shaded regions represents the inefficiently mixed region, from 550-$700\, \mathrm{km}$, see Section~\ref{subsec:conv_struct}.
            	The vertical black lines indicate the location of the Urca shell for the given pair. 
            }
        \end{figure*}

        % talk about how Urca are distributed. Urca shell locale etc.
        Once the simulation has relaxed, the three Urca pairs are distributed in the convection zone based on the location of their respective Urca shell and the pair's relation to the central carbon burning, see Figure~\ref{fig:urca_comp_profs}.
        In our reaction network, the A=21 pair is only loosely connected to the carbon burning via the $\isotm{O}{16}(n,\gamma)\isotm{O}{17}(\alpha, \gamma)\isotm{Ne}{21}$ pathway.
        In contrast, the A=23 pair is directly linked via the $\isotm{C}{12}(\isotm{C}{12},p)\isotm{Na}{23}$ reaction, one of the primary carbon burning rates.
        This explains why even in the NB simulation, with no beta-decays, there is still an appreciable amount of \isot{Na}{23} in the convection zone.
        Lastly, the A=25 pair is decoupled from the other rates in our network, so there is no increase in abundance of the Urca pair during the simulation.

        The location of the three Urca shells is also important to determining the distribution of each isotope in the Urca pair.
        The A=21 Urca shell is the furthest interior at $R \approx 170 \, \mathrm{km}$, which gives much more room for beta-decays compared to electron captures.
		The A=23 and A=25 Urca shells are located further out closer to the convective boundary at $R \approx 415 \, \mathrm{km}$ and  $R \approx 490 \, \mathrm{km}$.
        This leaves more room for electron captures than beta-decays, which is reflected in the ratio between the two isotopes for the A=23 and A=25 pairs.
        Another point to note is the relative abundance of each pairing.
        This is largely set by our initial model, see Table~\ref{tab:init_comp}, though there is also a net increase, particularly to the A=23 pair, due to the carbon burning.
        The A=23 pair is the more abundant by about an order of magnitude, while the other two pairings are roughly equivalent.

\begin{figure}
            \centering
            \includegraphics[width=\linewidth]{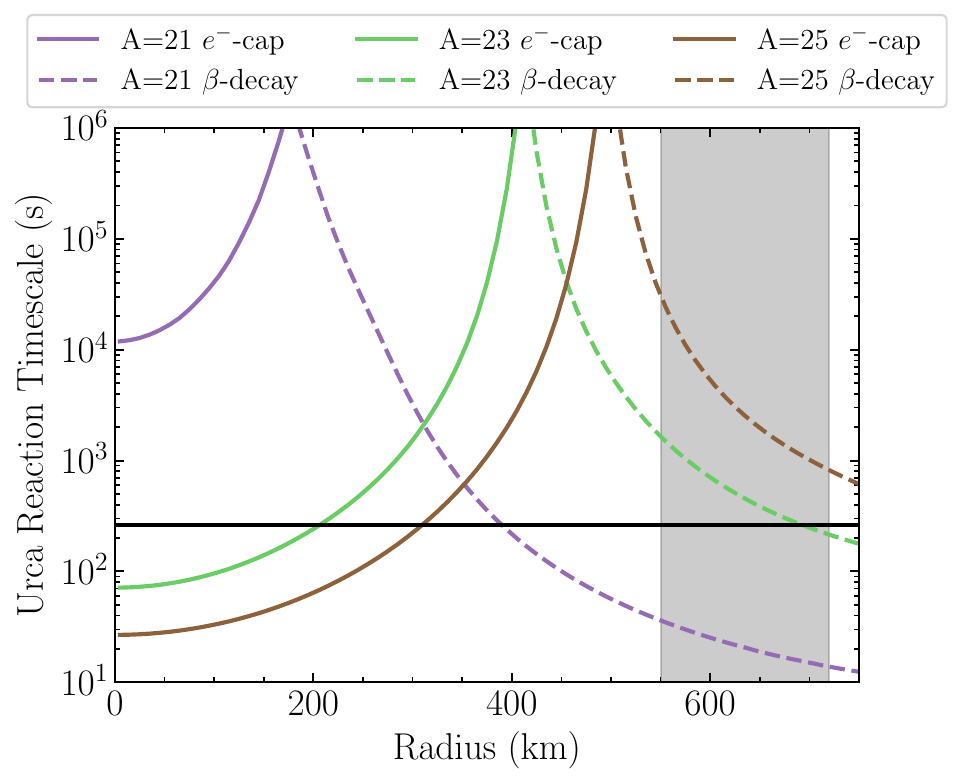}
            \caption{\label{fig:urca_tscale} 
				The reaction timescales for the A=21 (purple), A=23 (green), and A=25 (brown) Urca pairs vs.\ radial bin.
				The solid curves represent the electron capture, while the dashed curves represent the beta-decay.
                The horizontal black line marks the approximate the convective turnover time (${\sim}250$ s).
				The grey vertical shaded regions represents the inefficiently mixed region, from 550-$700\, \mathrm{km}$, see Section~\ref{subsec:conv_struct}.
            }
        \end{figure}

        % retrying to write this out.
        In the A=23 and A=25 case, the location of the two Urca shells lead to a larger abundance of electron capture products, i.e.\ \isot{Ne}{23} and \isot{Na}{25}.
        Still, there is a substantial region on the outer edge of the convection zone dominated by beta-decays. 
        In this region, there is a transition from primarily electron capture products to beta-decay products.
        This transition does not occur right at the Urca shell itself, but instead around $150 \, \mathrm{km}$ outside the respective Urca shells.
        At the shell, the Urca reactions are slow enough that the convective mixing largely sets the composition.
		The Urca reactions only become quick, relative to the convective mixing, around 100-$150 \, \mathrm{km}$ away from the Urca shell, see Figure~\ref{fig:urca_tscale}.
        The timescales shown are the average reaction timescale at a given radius and the convective turnover time, $\Tauconvm \approx 250 \, \mathrm{s}$.
        Through much of the convection zone, the turnover timescale is shorter than the Urca reactions, which leads to a more uniform distribution of the Urca pairs in most regions.

		Note, the convective timescale we compare the Urca reactions to is an approximation for the whole convection zone. 
		As shown in section \ref{subsec:conv_struct}, there is a less efficiently mixed region near the convective boundary from around 550-$700\, \mathrm{km}$.
		Here, the appropriate mixing timescale is likely longer than the ${\sim}250 \, \mathrm{s}$ we plot in Figure~\ref{fig:urca_tscale}.
		Because the mixing timescale is longer, the Urca beta-decays are likely just as important as the mixing in most of the 550-$700\, \mathrm{km}$ region. 
		The picture is further complicated, however, as it is these beta-decays that contribute to the compositional gradients that help slow down the convective mixing in this region.
		We discuss this situation in more detail in Section~\ref{sec:discussion}.

\begin{figure}
            \centering
            \includegraphics[width=\linewidth]{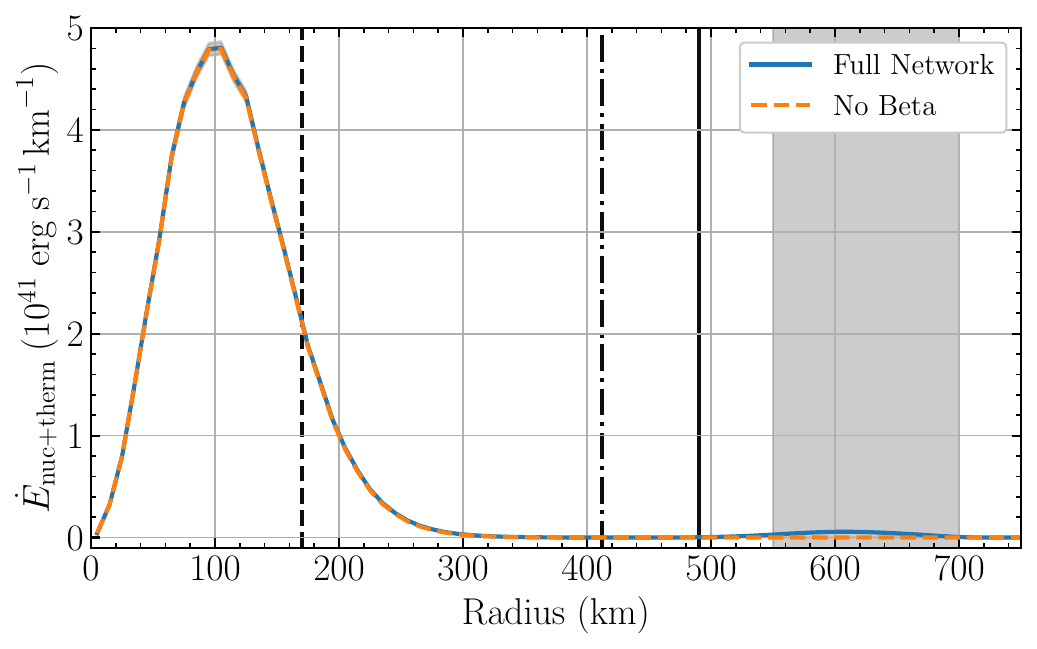}
            \caption{\label{fig:nuc_energy_profs} 
				The total energy generation rate from nuclear reactions per radial bin vs.\ radial bin.
				The blue curves represent the FN simulation, while the orange represent the NB simulation.
				Each curve represents the time averaged value for the given simulation over the last ${\sim}600 \, \mathrm{s}$, see Figure~\ref{fig:cburn_rate}.
				The grey vertical shaded regions represents the inefficiently mixed region, from 550-$700\, \mathrm{km}$, see Section~\ref{subsec:conv_struct}.
            	The vertical black lines indicate the location of the A=21 (dashed), A=23 (dot-dashed), and A=25 (solid) Urca shells.
            }
        \end{figure}

        % talk a lil about energy concerns I think, but want to frame in Urca relate.
        % optionally, move energy to like steady state calc where I talk about cburn and show that the central energy output are about the same which is what I'm tryign to get at.
        We calculate the total energy generation rate from nuclear reactions, including neutrino loss, and thermal losses due to adding/removing an electron.
        Due to the highly degenerate conditions in the white dwarf, there is a relatively large energy loss/gain associated with adding or removing an electron from the Fermi sea.
        A more detailed analysis of the thermal properties of the Urca reactions can be found in the following sources, \citet{barkat_wheeler1990, bruenn1973}; PA2026.
		In Figure \ref{fig:nuc_energy_profs}, we plot the profile of the total energy for the FN and NB simulations.
		Near the center, the energy is dominated by carbon burning, with the addition of some electron capture reactions.
		Since we keep the electron capture rates in the NB simulation, the two profiles match well inside about $500  \, \mathrm{km}$.
		Outside $500 \, \mathrm{km}$, we can see a small amount of energy generated by beta-decays in the FN simulation, primarily from the A=23 pair, which is the most abundant pair.
        Compared to the central carbon burning, this outer regions generates far less energy by more than two orders of magnitude.
		However, even though the energy contributions are modest, the convective Urca process still plays a significant role in setting the convective structure, as seen in Section~\ref{subsec:conv_struct}.

        % these last couple sentences might be a lil out of place and are more musings than anything lol. do kinda like throwing it out there tho.

    % maybe don't need sperate comparing section and just do that kinda throughout and hten leave hard core comparisons to the discussion section.
	\subsection{Comparing to Poca-Amor\'{o}s et al.\ 2026}\label{subsec:comp_f}
		We compare to the previous work in PA2026, which simulated the simmering phase with a different reaction network and different initial conditions.
		The reaction network used in PA2026 incorporated a simplified carbon burning network with only the A=23 Urca pair included.
		In comparison, the network presented in Section~\ref{sec:network} more accurately tracks the energy generation rate and net neutronization of the carbon burning.
		We focus this comparison on the FN1 simulation presented in PA2026, which we denote as the PA2026 simulation for clarity.
		The PA2026 initial model had a larger isentropic region (see Figure~\ref{fig:init_Tprofs}) and greater A=23 abundance, $5 \times 10^{-4}$.
		Otherwise, the central temperature and density are equivalent to the FN simulation.

\begin{figure}
            \centering
            \includegraphics[width=\linewidth]{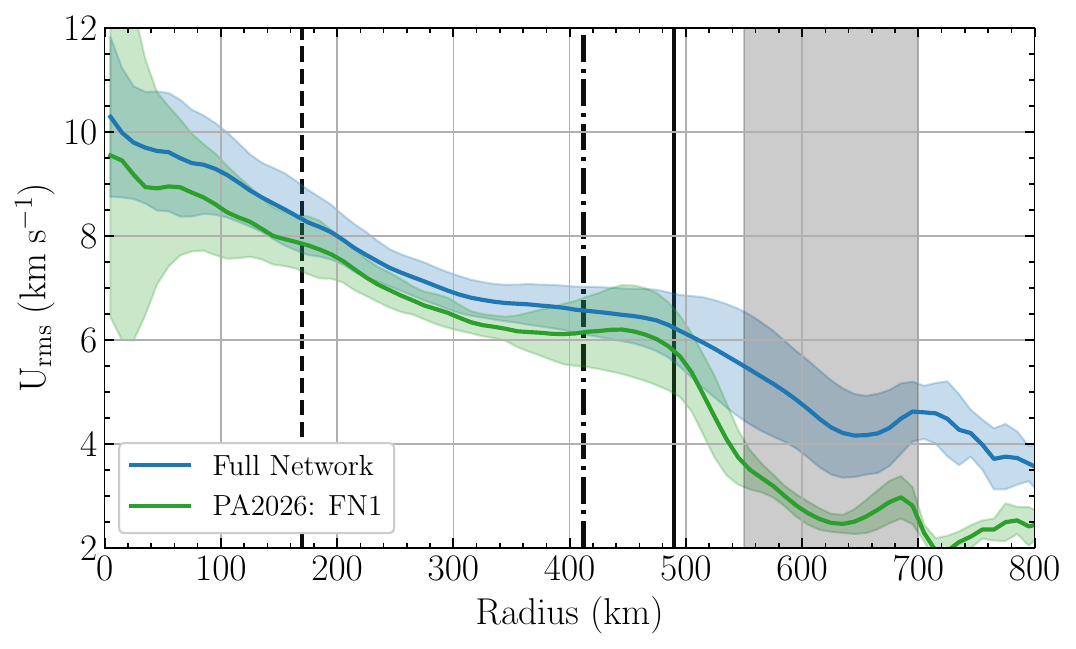}
            \caption{\label{fig:f_comp_vrms} 
		        The rms velocity profiles vs.\ binned radius. 
				The blue curve represents the time averaged value for the FN simulation over the last ${\sim}600 \, \mathrm{s}$, see Figure~\ref{fig:cburn_rate}.
                The green curve is from the PA2026 FN1 simulation with a similar time averaging over the last ${\sim}600 \, \mathrm{s}$ of the simulation.
                The shaded regions bounding each curve represents a single standard deviation away from the mean value.
				The grey vertical shaded regions represents the inefficiently mixed region, from 550-$700 \, \mathrm{km}$, see Section~\ref{subsec:conv_struct}.
            	The vertical black lines indicate the location of the A=21 (dashed), A=23 (dot-dashed), and A=25 (solid) Urca shells.
                }
        \end{figure}

		The convective structure of the FN and PA2026 simulations have some commonalities but are ultimately quite distinct.
		The FN simulation, as described in section~\ref{subsec:conv_struct}, has an inefficiently mixed region on the edge of the convection zone, around 550-$700\, \mathrm{km}$.
		In the PA2026 simulation, the convection zone is largely restricted to a radius of about $540 \, \mathrm{km}$, with an outer isentropic region that extends out to nearly $700\, \mathrm{km}$.
		This outer region of the PA2026 simulation indicates the possibility of a split convection zone structure, that is two distinct convection zones connected by an intermediate region that is stabilized to convection due to compositional gradients.
		We can see the difference in the convective structure of the two simulations in Figure~\ref{fig:f_comp_vrms}.
		Here, we see similar convective velocities inside about $500 \, \mathrm{km}$. 
		Outside this radius, the FN simulation slowly declines while the PA2026 simulation much more quickly declines to a lower velocity.
		
\begin{figure}
            \centering
            \includegraphics[width=\linewidth]{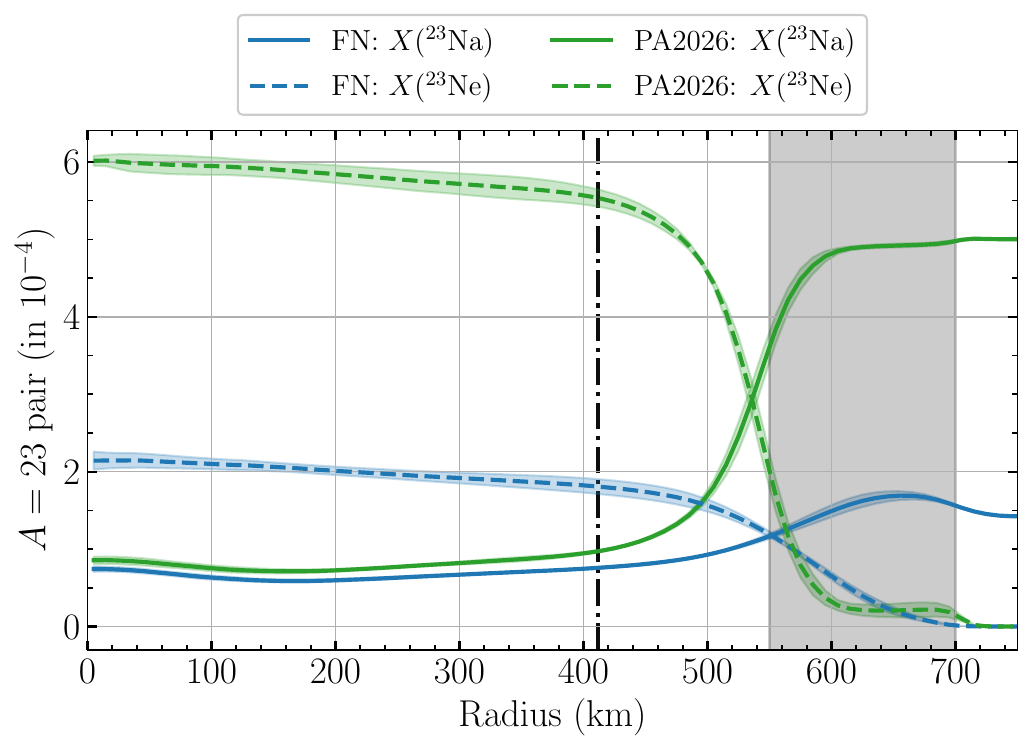}
            \caption{\label{fig:f_comp_urca23} 
                Density-weighted angle-averaged composition profiles for the A=23 Urca pair vs.\ binned radius. 
				The blue curve represents the time averaged value for the FN simulation over the last ${\sim}600 \, \mathrm{s}$, see Figure~\ref{fig:cburn_rate}.
                The green curve is from the PA2026 FN1 simulation with a similar time averaging over the last ${\sim}600 \, \mathrm{s}$ of the simulation.
                The shaded regions bounding each curve represents a single standard deviation away from the mean value.
				The grey vertical shaded regions represents the inefficiently mixed region, from 550-$700\, \mathrm{km}$, see Section~\ref{subsec:conv_struct}.
            	The vertical black lines indicate the location of the A=21 (dashed), A=23 (dot-dashed), and A=25 (solid) Urca shells.
                }
        \end{figure}

		An explanation for why convective Urca only slows mixing in the FN simulation, while it largely prevents mixing in the PA2026 simulation is the difference in A=23 Urca abundance.
		This Urca pair is expected to be the most important and has the highest mass fraction as seen in Figure~\ref{fig:urca_comp_profs}.
		Comparing the two simulations, see Figure~\ref{fig:f_comp_urca23}, we see the PA2026 has roughly three times the abundance.
		Besides this difference in magnitude, the distributions share a similar structure. 
		That is, the point where \Xisot{Na}{23} and \Xisot{Ne}{23} are equal occurs at about $550 \, \mathrm{km}$ in both simulations.
 
\begin{figure}
            \centering
            \includegraphics[width=\linewidth]{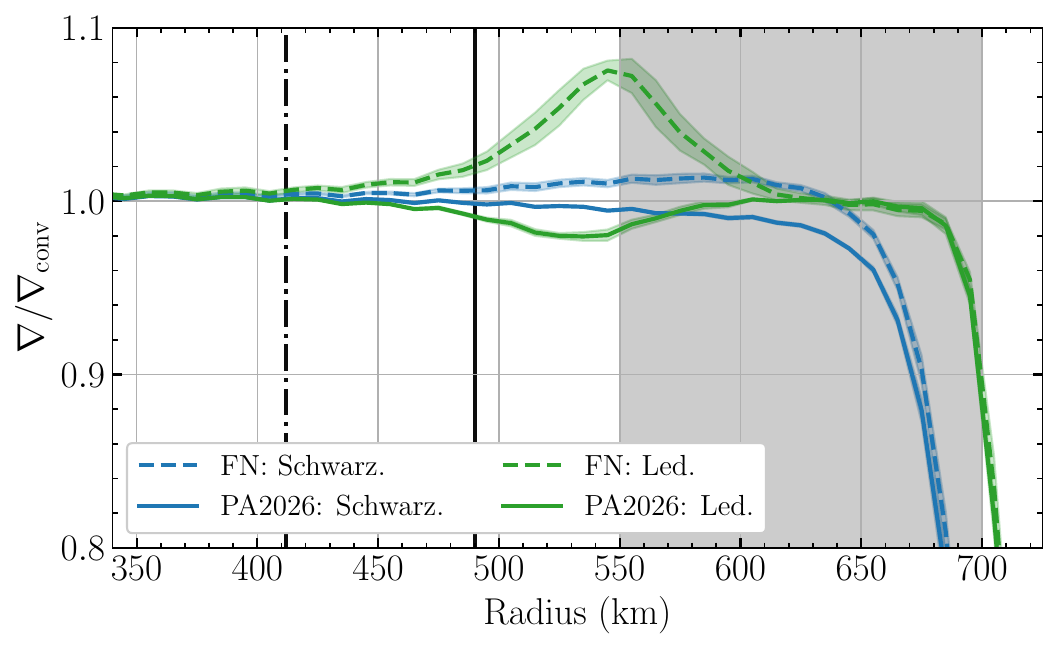}
            \caption{\label{fig:f_comp_conv} 
                Density-weighted angle-averaged profiles of the ratio of convective gradients vs.\ binned radius. 
				The dashed curves relate to the Schwarzschild criterion $\nabla / \nabla_{\mathrm{ad}}$.
				The solid curves relate to the Ledoux criterion $\nabla / \nabla_{\mathrm{Led}}$.
				The blue curve represents the time averaged value for the FN simulation over the last ${\sim}600 \, \mathrm{s}$, see Figure~\ref{fig:cburn_rate}.
                The green curve is from the PA2026 FN1 simulation with a similar time averaging over the last ${\sim}600 \, \mathrm{s}$ of the simulation.
                The shaded regions bounding each curve represents a single standard deviation away from the mean value.
				The grey vertical shaded regions represents the inefficiently mixed region, from 550-$700\, \mathrm{km}$, see Section~\ref{subsec:conv_struct}.
            	The vertical black lines indicate the location of the A=23 (dot-dashed) and A=25 (solid) Urca shells.
                }
        \end{figure}

		We also compare the differences in the convective stability between the two simulations in Figure~\ref{fig:f_comp_conv}.
		This demonstrates the stark difference between the simulations as PA2026 has an apparent split convection zone.
		Though both simulations have an apparent semi-convective region (Schwarzschild unstable but Ledoux stable), the gradient in the FN simulation remains stable according to the Ledoux
criterion in the region outside $500 \, \mathrm{km}$.  
		This differs from the PA2026 simulation, where the $\nabla/\nabla_\mathrm{Led}$ ratio rises back up to one, indicating another Ledoux-unstable region at the outer edge.
		Of note, the PA2026 shows evidence of larger stabilizing compositional gradients as compared to the FN simulation.  
		Lastly, the difference in the initial temperature profile can be seen in where the hard convective boundary is located in each simulation, i.e.\ where the ratio begins dipping well below one.
		There is about a $50 \, \mathrm{km}$ difference that accounts for roughly half of the PA2026 simulation's outer convectively unstable region.
		
		The entropy profiles of the two simulations, shown in Figure~\ref{fig:entropy}, further demonstrate the key differences in the structure of the two convection zones.
		The difference in the central entropy between the two simulations is due to differences in temperature and composition, which still largely reflects the initial composition seen in Table~\ref{tab:init_comp}.
		%In particular, the FN simulation started with with less \isot{O}{16}, and less of the A=23 Urca pair.
		For each simulation, a dip in the entropy coincides with the compositional gradient formed by the A=23 Urca pair.
		The PA2026 simulation has a much larger dip in terms of magnitude, again likely due to abundance, that leads to a flat entropy profile spanning from 600-$700\, \mathrm{km}$.
		In contrast, the FN simulation shows a dip that reaches a minima and then increases into the hard entropy boundary from our initial model.

		We have a mild concern that, compared to the structure here, the initial
		condition used in the PA2026 simulation may have an initial isentropic
		region that is too large, leading to the second exterior
		convection zone.  The problem arises from the uncertainty in the size, in
		both magnitude and radial extent, that the entropy ``dip" (created by the
		super-adiabatic gradient in the semi-convective region) should occupy.  Using
		Figure 11 as a reference, it may be that the flat entropy region, between 600-$700\, \mathrm{km}$ in radius, in the PA2026 simulation is a remnant of the initial condition.  
		That is, if the model had been evolved from earlier time, and thus heated from the inside out, that flat region would be at a lower temperature, and therefore
		lower entropy. 
		The resulting entropy profile would have a more pronounced dip but no flat region to become an outer convection zone.  
		To properly reflect and investigate this entropy structure, future work should develop a consistent way to include a semi-convective region in
		the initial condition.  Notably, based on the 3D simulations performed so
		far, such a region should be placed outside the Urca shell proper, an outward
		shift from where it might have been placed based on previous 1D work (e.g.\ \citealt{piersanti2022}).

\section{Discussion}\label{sec:discussion}
		The main advancement of the presented simulations is the larger more comprehensive reaction network.
		This network accurately models the carbon burning during this stage of simmering and includes all relevant Urca pairs.
		The inclusion of this much larger reaction network was computationally feasible due to GPU acceleration and the reduced cost of establishing a reasonable convective field without evolving the base state, see Section~\ref{subsec:init_evol}.
		In addition, setting reasonable initial Urca distributions based on a quick mixing limit aided in quickly realizing a steady state~\citep{boydproceedings}.
		These principles will help in future work as we explore more models with varying initial conditions (e.g.\ central temperature, density, convective boundary, etc.).

		Similar to the work in PA2026, we ran two simulations, with (FN) and without (NB) the convective Urca process.
		In comparing the FN and NB simulation, we found that the extent of mixing and convective velocities were similar, see Figure~\ref{fig:mconv_vrms}.
		The major difference between the simulations was in the region near the convective boundary from about 550-$700\, \mathrm{km}$, as highlighted in Figures~\ref{fig:ne20_eta_profs} and~\ref{fig:vrms_profs}.
		In this region, the convective velocities and efficiency of mixing drop significantly for the FN simulation, demonstrating the hindering effects of the convective Urca process.
		This region follows a super-adiabatic temperature gradient that is stabilized against convection by compositional gradients, see Figure~\ref{fig:conv_grad}.
		It is important to note that convective mixing still extends to the same radius for both simulations.
		In the FN simulation, the convective Urca process only slows convection in the outer region, it does not restrict convection to a smaller radius, as in some previous work (\citealt{stein-wheeler2006}; PA2026).

		Although the FN simulation includes the A=21 and A=25 Urca pairs, the A=23 Urca pair is the most important to the convective Urca process.
		The A=23 pair is roughly an order of magnitude more abundant than the other two Urca pairs, and is a direct product of the carbon burning.
		We find that the inefficiently mixed region is located outside both the A=23 and A=25 Urca shells, but when looking at the temperature profile, we can see a super-adiabatic temperature gradient forming inward to the A=25 Urca shell.
		Additionally, the A=23 beta-decay timescale approaches the convective mixing timescale in the inefficiently mixed region (see Figure~\ref{fig:urca_tscale}), again indicating the importance of this Urca pair.
		Compared to the location of the A=23 Urca shell, the A=21 and A=25 Urca shells are not well positioned to produce separate regions in the convection zone where electron capture and beta-decay can each be active on a timescale similar to convective mixing.
		The A=21 Urca shell is close to the center, and thus the electron capture timescale remains quite long, over ${\sim} 10^4 \, \mathrm{s}$.
		The A=25 Urca shell is close to the convective boundary and so the beta-decay timescale only dips below ${\sim} 10^3 \, \mathrm{s}$ right at the edge of the convection zone ($\Rconv \approx 725 \, \mathrm{km}$).
		Without a reasonably large electron capture region and beta-decay region, the A=21 and A=25 Urca pairs cannot have as significant an effect as the A=23 Urca pair.

		An interesting point of comparison for the FN simulation is to the simulations from PA2026, particularly the FN1 model, which demonstrated the restrictive nature of the convective Urca process.
		%The FN simulation and PA2026 simulation started with similarly sized initial isentropic regions, 0.9~\Msun and 1~\Msun, see Figure~\ref{fig:init_Tprofs}.
		In PA2026, despite an initial isentropic region of 1~\Msun, the convective mixing was largely contained to a significantly smaller region ($\Mconv \approx 0.66 \, \Msun$), with a potential outer convection zone split away from the main core convection zone.
		There are multiple differences between the FN and PA2026 simulations that could explain the difference in resulting convective structures.
		For one, PA2026 implemented a far reduced carbon burning network. 
		However, the convective velocities in the interior of the convection zone for each simulation are comparable (see Figure~\ref{fig:f_comp_vrms}).
		This indicates the differences in carbon burning is not the primary driver of differences in the FN and PA2026 convection zones.
		Instead, we believe key differences in the initial progenitor model plays a more important role.

		The two key differences in the initial models is the abundance of the A=23 Urca pair (see Figure~\ref{fig:init_urca}) and the size of the isentropic region (see Figure~\ref{fig:init_Tprofs}).
		The PA2026 simulation began with nearly three times the A=23 Urca abundance as the FN simulation.
		This added abundance likely enhances the effects of the convective Urca process, leading to more reactions and larger compositional gradients.
		The difference in initial isentropic regions may also be important as a larger isentropic region may provide ample room for a split convection zone structure to form.
		In a split structure, there are two separate regions, an inner and outer convection zone, in which the abundance and the entropy are nearly uniform.
		These convection zones are separated by an intermediate shell region where the temperature gradient is significantly steeper than adiabatic but with a counter-balancing
gradient in the neutron excess, which prevents straightforward convection.  
		There may be overshoot-like mixing across this intermediate shell region, leading to some coupling between the two convection zones.
		Because the hard entropy barrier (at $R=700 \, \mathrm{km}$ in Figure~\ref{fig:entropy}) is closer to the Urca shell in the FN simulation, there is little room for a secondary isentropic region (i.e.\ outer convection zone) to form, as seen in the PA2026 simulation (see green curve flattens out from 600-$700\, \mathrm{km}$ in Figure~\ref{fig:entropy}).

		Setting aside the differences in composition, it is important to note that the FN and PA2026 are simulating fundamentally different progenitors because the isothermal, outer temperature is different.
		The isothermal temperature profile is determined by prior evolution of the white dwarf star, including the cooling and accretion histories prior to the simmering phase.
		This suggests the convective Urca process may be somewhat sensitive to these conditions, which we would expect to vary over the population of progenitors.

		An important note concerning the FN simulation, and to a lesser extent the PA2026 simulation, is the relatively small abundance of the A=23 Urca pair.
		Our initial composition is based on the start of the simmering phase.
		However, the central temperature, ${\sim}5.5\times10^8$ K, aligns closer to the latter end of the simmering phase, and thus there should be an enhancement of the Urca pairs due to carbon burning, particularly for the A=23 pair which should have an abundance closer to $10^{-3}$.
		Along with this enhancement, the Urca abundances are also sensitive to the metallicity of the white dwarf. 
		Overall, the FN simulation represent a relatively conservative picture of the convective Urca process and may align more closely with a lower metallicity progenitor model.

		A difficulty in constructing proper initial models for these 3D simulations is properly accounting for the prior evolution during the simmering phase.
		Our general understanding of the simmering phase is that the central temperature will rise due to carbon burning, creating a temperature gradient that drives convection.
		As the temperature steadily rises, the convection zone will grow to encompass more and more of the white dwarf.
		Eventually, the convection boundary will meet an Urca shell, likely A=21 or A=23 depending on the central density of the white dwarf.
		At this Urca shell, there will be a pre-existing compositional gradient that will impede the further growth of the convection zone.
		What our presented simulations still do not answer is whether the convection zone will continue to grow past the Urca shell~\citep[e.g.][]{martinez-rodriguez2016}, or whether the compositional gradients and convective Urca process will restrict the size of the convection zone to the Urca shell~\citep[e.g.][]{piersanti2022}.

		Important to this discussion is that the present simulations, as well as the results in PA2026, show that mixing is not restricted at the A=23 Urca shell, but instead further out around ${\sim}150 \, \mathrm{km}$ past the shell.
		Here, the beta-decay timescale drops to under 1000 seconds, or of order a few convective turnover times (see Figure~\ref{fig:urca_tscale}).
		Because the Urca reactions are very slow at the Urca shell, it may be the case that even at earlier times, when convection is slower, mixing must extend meaningfully past the Urca shell for the convective Urca process to hinder or restrict that convective mixing.
		The relative importance of overshoot of convective motion (regardless of the composition gradient) compared to the counter-buoyant tendency of the Urca products is also unclear.
		Even if the mixing is confined near the Urca shell for some period during the growth of the convection zone, the overshoot
		boundary of this confined region may extend well past the Urca shell due to the relative slowness of the Urca process near the boundary.

		A related question is how the convective boundary will manifest under these conditions.
		If the convection zone is restricted or hindered around the Urca shell, a super-adiabatic temperature gradient will form to connect the convective core to the stable envelope.
		As the carbon burning continues to heat the convection zone, this temperature gradient at the convective boundary will grow steeper as well.
		How long the convective Urca process, and other compositional gradients, can stabilize this region is still unclear.
		
		Something pertinent to understanding the convective Urca process is the extent to which compositional gradients alone can explain the hindering of convective mixing.
		As discussed in Section \ref{sec:urca}, there are a number of ways the convective Urca process can reduce the efficiency of convective mixing, e.g.\ altering the buoyancy of the fluid and transporting electrons up a chemical gradient. 
		A well implemented model of the convective Urca process should account for these effects. 
		At the same time, the compositional gradients formed by the Urca process appear to play a significant role in these simulations and may be the most important piece to accurately modeling the convective Urca process.
		Properly quantifying the relative importance of these different aspects of the convective Urca process should help inform a convective Urca model that can be implemented in 1D stellar evolution models.

\section{Conclusions and Future Work}\label{sec:conclusion}
	We conducted simulations of the convective Urca process with a comprehensive carbon burning network and three Urca pairs.
	Considering the effects of the convective Urca process, we find an inefficiently mixed region, as compared to the rest of the convection zone, outside the A=23 and A=25 Urca shells.
	This region is not present when we ignore beta-decays, and points to the hindering effect the convective Urca process has on convection.
	Despite this hindering effect, mixing still extends to the edge of our initial isentropic region in both the FN and NB simulation.
	This result sits in contrast to previous multi-D simulation work, which found convection was largely contained to smaller radii (\citealt{stein-wheeler2006}; PA2026).
	The difference between these results may be due to differences in the size of the isentropic region and/or in the abundance of the A=23 Urca pair.

	In this work, we found the A=23 Urca pair had the largest effect on the convective structure.
	This is in part due to the greater abundance of the A=23 Urca pair, as compared to the A=21 and A=25 pairs, as well as the location of the A=23 Urca shell, which allowed for significant electron capture and beta-decay regions.
	The inefficiently mixed region is located about $150 \, \mathrm{km}$ outside the A=23 Urca shell, and corresponds to the region in which the A=23 beta-decay timescale approaches the convective mixing timescale.

	From this work, and previous work (B2025; PA2026), it is clear that the initial model plays a significant factor in these simulations.
	The prior evolution during the simmering phase will set both the temperature profile and compositional profiles.
	In future work, we wish to explore this space in applying larger compositional gradients that would better reflect the amount of prior carbon burning and the extent of convective mixing.
	In addition, this carbon burning produces the A=23 pair, and to a lesser extent the A=21 pair. 
	As such, we wish to explore the effects of greater Urca abundance and the sensitivity of the convective Urca process to changes in that abundance.
	The restrictive nature of the convective Urca process likely will generate regions near a convective boundary where the temperature profile is steeper than the adiabatic.
	This can be seen in the FN simulation even during the relatively short simulation time.
	As such, a proper initial model should account for these features and is an aspect we wish to explore in more detail.
	Lastly, we will explore different central temperatures, particularly lower temperatures that correspond to the earlier simmering phase.

\section*{Acknowledgments}
    Visualizations and part of this analysis made use of \yt\ \cite{turk2011}.
    The reaction networks were generated using the \pynucastro\ library \cite{smith2023}.
	This research has made use of NASA’s Astrophysics Data System Bibliographic Services.
    This research was supported in part by the US Department of Energy (DOE) under grant DE-FG02-87ER40317.
	D.M.T. acknowledges support from the National Science Foundation under grant No. 2307442.
    This research used resources of the National Energy Research Scientific Computing Center, which is supported by the Office of Science of the U.S. Department of Energy under Contract No. DE-AC02-05CH11231, using NERSC award NP-ERCAP0027167.
    The authors would like to thank Stony Brook Research Computing and Cyberinfrastructure and the Institute for Advanced Computational Science at Stony Brook University for access to the high-performance SeaWulf computing system, which was made possible by \$1.85M in grants from the National Science Foundation (awards 1531492 and 2215987) and matching funds from the the Empire State Development’s Division of Science, Technology and Innovation (NYSTAR) program (contract C210148).

\textit{Software}: {\amrex\ \citep{zhang2019},                 
          {\sffamily GNU Parallel} \citep{gnu_parallel},
          \maestro\ \citep{fan2019,maestroex_joss},
          {\sffamily matplotlib}\  \citep{Hunter:2007},
          {\sffamily NumPy}\ \citep{numpy2020},
          {\sffamily pandas} \citep{pandas},
          \pynucastro\ \citep{pynucastro,smith2023},
          {\sffamily SymPy} \citep{sympy}, yt \citep{turk2011}}
         
\bibliographystyle{aasjournal}
\bibliography{urca.bib}{}

\end{document}